\documentclass[a4paper,11pt]{article}
\usepackage{jcappub} % for details on the use of the package, please see the JINST-author-manual
\usepackage{lineno}
\usepackage[dvipsnames]{xcolor}

\usepackage{mathtools}

\usepackage{slashed}
\usepackage{cleveref}
\crefname{figure}{figure}{figures} 
\usepackage[normalem]{ulem}
\usepackage{multirow}
\usepackage{comment}

\usepackage{tablefootnote} 

\usepackage{cancel}

\renewcommand{\d}{\mathrm{d}}
\newcommand{\be}{\begin{equation}}
\newcommand{\ee}{\end{equation}}

\arxivnumber{2511.15790}
\title{\boldmath How light can ALP dark matter be?}

\author{Kierthika Chathirathas}
\author{and Thomas Schwetz}
\affiliation{Institute for Astroparticle Physics (IAP), Karlsruhe Institute of Technology (KIT),\\ Hermann-von-Helmholtz-Platz 1, 76344 Eggenstein-Leopoldshafen, Germany}

\emailAdd{kierthika.chathirathas@kit.edu}
\emailAdd{schwetz@kit.edu}

\abstract{We assume axion-like particles (ALPs) to provide the full dark matter abundance and derive various lower bounds on the ALP mass. 
We contrast the post- and pre-inflationary symmetry breaking cases and present allowed regions in the plane of ALP mass and energy scale of inflation. 
For the post-inflationary case, we revisit bounds from isocurvature perturbations taking into account that, as suggested by simulations, axion radiation by cosmic strings during the scaling regime provides the dominant production mechanism of dark matter, obtaining significantly weaker limits than previously. Combining isocurvature, with constraints from black hole superradiance and free streaming, we find that the bound $m_a \gtrsim 10^{-17}$~eV applies for most cases considered here. It can be potentially relaxed to $\sim 6\times 10^{-19}$~eV only in the post-inflationary case with a strongly temperature-dependent axion mass, subject to uncertainties on the axion emission spectrum. Significantly stronger bounds are obtained in the post-inflationary scenario from the non-observation of CMB tensor modes, which can be as strong as $m_a > 5\times 10^{-7}$~eV for small reheating efficiencies, $\epsilon \lesssim 5\times 10^{-4}$.}

\begin{document}
\maketitle
\flushbottom

%\linenumbers
%%%%%%%%%%%%%%%%%%%%%%%%%%%%%%%%%%%%%%%%%%%%%%%%%
\section{Introduction}
%%%%%%%%%%%%%%%%%%%%%%%%%%%%%%%%%%%%%%%%%%%%%%%%%

Axion-like particles (ALPs) have gained significant attention in recent years due to their potential to address unresolved problems in particle physics and cosmology. The original QCD axion emerged from the Peccei-Quinn mechanism \cite{Peccei:1977hh,Peccei:1977ur}, which was proposed to solve the strong CP problem in quantum chromodynamics \cite{Weinberg:1977ma,Wilczek:1977pj}. In this work, we focus on a broader class of light, very weakly interacting pseudo-scalar particles. These ALPs naturally arise in many extensions of the Standard Model and string theories as pseudo-Nambu-Goldstone bosons resulting from the spontaneous breaking of a global $U(1)$ symmetry at an energy scale $f_a$, commonly known as the axion decay constant. Our main motivation in this work is the fact that such particles provide an attractive candidate for dark matter in the universe, see, e.g., \cite{Sikivie:2006ni,Marsh:2015xka} for reviews. In particular, ultra-light ALP dark matter, broadly in the range of $10^{-22}$ to $10^{-18}$~eV, is often denoted as ``fuzzy dark matter'' and leads to a multitude of cosmological effects \cite{Hu_2000, Hui:2016ltb}, see Ref.~\cite{Eberhardt:2025caq} for a recent review.

The cosmological evolution of ALPs depends strongly on the time of the global $U(1)$ symmetry breaking relative to the end of inflation. In the pre-inflationary scenario, the symmetry is broken before or during inflation and is never restored afterwards. 
A single causally connected patch is inflated to the entire observable universe, which leads to a homogeneous initial axion field value, usually parametrized by the misalignment angle. The corresponding axion dark matter emerges when the axion mass becomes comparable to the Hubble expansion rate and the field starts oscillating. The resulting dark matter energy density depends sensitively on the initial misalignment angle \cite{Preskill:1982cy,Abbott:1982af,Dine:1982ah,Turner:1985si}.

In the post-inflationary scenario, the symmetry is broken after inflation. The axion field assumes random initial values in causally disconnected regions, each limited in size by the horizon at the time of symmetry breaking. Topological defects, in particular a network of cosmic strings, can form during the phase transition \cite{Kibble:1976sj}. During the cosmic expansion, this system will tend towards a scaling configuration, keeping the average number of strings per Hubble volume roughly of order one, up to logarithmic violations. In order to maintain such a configuration, energy has to be released, which happens via the formation and subsequent collapse of closed string loops predominantly into relativistic axions \cite{Davis:1986xc}. Indeed, recent results from numerical simulations suggest, that these axions generated by the string network (once redshifted to non-relativistic momenta) provide the dominant contribution to the axion dark matter abundance in the post-inflationary scenario \cite{Gorghetto2018,Gorghetto2021,Buschmann2022,Kim2024,Saikawa2024,Benabou2025,Kim:2024dtq}, see also \cite{Klaer:2017ond,Klaer:2019fxc,Kawasaki:2018bzv,Pierobon:2023ozb,Correia:2024cpk}.

In this work, we revisit the constraints from isocurvature perturbations in the post-inflationary ALP dark matter scenario \cite{Feix2019,Feix2020,Irsic2020}. These previous analyses assumed axion dark matter production via the misalignment mechanism, which we revise by using the dominating contribution from string network radiation, see also \cite{GorghettoGW2021}. In particular, we apply bounds on the isocurvature-to-adiabatic power spectra ratio from Lyman-$\alpha$ observations~\cite{Irsic2020} as well as from ultra-faint dwarf galaxies~\cite{Graham:2024hah} and derive 
updated lower bounds on the ALP mass, finding significantly weaker constraints than obtained in the above mentioned studies. We discuss the dependence of these bounds on ALP model properties as well as on theoretical uncertainties related to the axion production mechanism. 

Furthermore, we point out that the non-observation of tensor modes in the CMB allows us to place a complementary lower bound on the axion mass in the post-inflationary scenario, which potentially excludes large parts of the low-mass axion dark matter region, especially for inflation models with small reheating efficiencies. 

Finally, we contrast the post- and pre-inflationary ALP dark matter scenarios and derive allowed regions for either scenario in the plane of the axion mass and the energy scale of inflation. Combining bounds from isocurvature and CMB tensor modes with constraints from black hole superradiance \cite{Arvanitaki:2009fg}, we show that low-mass ALP dark matter in the fuzzy dark matter region is severely disfavoured. 

The outline of the paper is as follows. In \cref{sec:framework}, we introduce the general ALP dark matter model assumed here and provide a short overview of various bounds, focusing on the low-mass region. In \cref{sec:ALP_DM_post,sec:iso-post,sec:tensor-modes}, we discuss various aspects of the post-inflationary scenario: we review the dark matter production from string network radiation in \cref{sec:ALP_DM_post}, derive the bounds from isocurvature in \cref{sec:iso-post}, and introduce the constraint from CMB tensor modes in \cref{sec:tensor-modes}. \Cref{sec:pre} contains a brief review of the pre-inflationary scenario, including the dark matter production from the misalignment mechanism as well as isocurvature constraints. In \cref{sec:post-vs-pre}, we show allowed regions for post- and pre-inflationary cases in the plane of axion mass and energy scale of inflation, combining the various bounds discussed above with bounds from black hole superradiance. We summarize our findings in \cref{sec:conclusion}. Supplementary material is provided in \cref{sec:constant-string-density,sec:random-phase-model,app:self-int,app:iso_pre}.

Throughout this paper, we assume that ALPs provide all dark matter in the universe, with the normalized energy density of $\Omega_{\rm DM} h^2 = 0.12$ \cite{Planck:2018vyg}.
We adopt a standard flat FLRW cosmology, with a radiation-dominated universe after reheating and model the effective number of relativistic degrees of freedom for both, energy density and entropy density, as smooth, temperature-dependent functions fitted to lattice QCD and perturbative results, following the approach of~\cite{Wantz2010}. 

%%%%%%%%%%%%%%%%%%%%%%%%%%%%%%%%%%%%%%%%%%%%%%%%%
\section{Framework}\label{sec:framework}
%%%%%%%%%%%%%%%%%%%%%%%%%%%%%%%%%%%%%%%%%%%%%%%%%

For concreteness, and following the original Peccei-Quinn mechanism, we consider a complex scalar field $\phi$ described by the Lagrangian
\begin{equation}
    \mathcal{L}_\phi = \partial_\mu \phi^* \partial^\mu \phi - \lambda \left( |\phi|^2 - \frac{f_a^2}{2} \right)^2.
\end{equation}
The global $U(1)$ symmetry is spontaneously broken by the vacuum expectation value $f_a/\sqrt{2}$ and we
expand $\phi$ around its minimum as
\begin{equation}
    \phi(x) = \frac{f_a + r(x)}{\sqrt{2}} \, e^{i\theta(x)} \,.
\end{equation}
The heavy radial field $r(x)$, commonly referred to as \textit{saxion},  has a mass $m_r = \sqrt{2\lambda}f_a$.
For the sake of definiteness, we will assume $m_r = f_a$. The angular field $\theta(x)$ is defined as $\theta(x) = a(x)/f_a$, where $a(x)$ is the Goldstone boson, which is identified as the \textit{axion}. We assume that it obtains a potential at low temperature due to explicit breaking of the global $U(1)$. In analogy to QCD, we consider a potential of the form
\begin{equation}\label{eq:V}
  V(T, \,\theta) = f_a^2 m_a^2(T)(1-\cos{\theta)} \,,
\end{equation}
and parametrize the temperature-dependent axion mass as
\begin{equation}\label{eq:ma_T}
    m_a(T) = \min\left[ \frac{\Lambda^2}{f_a},\ \beta \frac{\Lambda^2}{f_a} \left( \frac{\Lambda}{T} \right)^{\alpha/2} \right]
    \quad \text{with} \quad  \Lambda^2 = m_a f_a \,.
\end{equation}
Note that whenever we use $m_a$ without explicitly including the temperature dependence, we refer to the zero-temperature axion mass. The parameter $\alpha$ controls how sharply the axion mass turns on as temperature decreases, whereas $\beta$ is a dimensionless coefficient that allows for the possibility that the zero-temperature mass may not be precisely reached at $T = \Lambda$ but is shifted to $T_c = \beta^{2/\alpha}\Lambda$. The temperature-independent case $m_a(T) = m_a$ is labeled by $\alpha=0$. We can achieve an evolution similar to the QCD axion by setting $\Lambda = 75.5 \text{ MeV}, \, \alpha = 8$ and $\beta = 10$~\cite{di_Cortona_2016, Borsanyi2016}. In our work, we treat $m_a$ and $f_a$ as independent parameters. For a given $m_a$ (and $\alpha,\,\beta$), we determine $f_a$ by requiring that axions provide all dark matter. 

The cosmological evolution of ALPs depends crucially on the order in which cosmic inflation and the spontaneous breaking of the global $U(1)$ symmetry occur. The case when spontaneous symmetry breaking occurs before the end of inflation and is not restored afterwards is called the \textit{pre-inflationary scenario}, whereas the opposite case is referred to as the \textit{post-inflationary scenario}. We define 
\begin{align} \label{eq:Tsym}
    T_{\rm sym}\equiv \max\left[ T_{\rm GH},\,T_{\rm max} \right] \,,
%    = \max\left[ \sqrt{\frac{8\pi}{3}}\frac{E_I^2}{M_{\rm pl}}, \, \epsilon E_I \right] \,
\end{align}
where $T_{\rm GH}$ is the Gibbons–Hawking temperature associated with the horizon during inflation \cite{Gibbons1977}:
\begin{equation}\label{eq:T_GH}
    T_\mathrm{GH} = \frac{H_I}{2\pi} \,,
\end{equation}
with $H_I$ denoting the Hubble parameter during inflation, and
\begin{equation}
    T_\mathrm{max} = \epsilon\, E_I 
\end{equation}
is the maximum thermalization temperature after inflation,
where $E_I$ denotes the energy scale of inflation and $0 < \epsilon < 1$ parametrizes the reheating efficiency. 
$H_I$ and $E_I$ are related through the Friedmann equation:
\begin{equation}\label{eq:HI_EI}
    H_I^2 = \frac{8\pi}{3 M_\mathrm{pl}^2} E_I^4 \,,
\end{equation}
where $M_{\rm pl} = 1.2\times 10^{19}$~GeV is the Planck mass. Hence, we can consider $T_{\rm sym}$ as a function of $E_I$ and $\epsilon$. For $E_I < \sqrt{3\pi/2}\,\epsilon M_{\rm pl}$, the temperature relevant for symmetry breaking is $T_{\rm max}$, while the Gibbons-Hawking temperature is relevant for large $E_I$ and sufficiently small $\epsilon$. The condition for the two ALP scenarios is then given by (see, e.g.,~\cite{Hertzberg:2008wr, Visinelli_2017, Visinelli_2009}),
\begin{equation}
    \label{eq:pre_post_condition}
    \begin{array}{l@{\qquad}l}
     f_a > T_{\rm sym} \,:   &  \text{pre-inflation} \,,\\[2mm]
     f_a < T_{\rm sym} \,:   &  \text{post-inflation} \,.
    \end{array}
\end{equation}

A general upper bound on $E_I$ is obtained from the non-observation of primordial tensor modes in the CMB. Planck and Bicep/Keck CMB observations place an upper bound on the tensor-to-scalar ratio of $r < 0.037$ at 95\%~CL, which implies~\cite{Campeti2022,BICEP:2021xfz,Planck:2018jri}
\begin{equation}\label{eq:EI_tensor_modes}
    E_I < 1.4\times 10^{16}\,{\rm GeV} \,.
\end{equation}
We discuss the implications of this bound in the post-inflationary case in \cref{sec:tensor-modes}. Future CMB missions may improve the bound on $r$ up to one order of magnitude \cite{Abazajian2022}. However, as $r$ enters only proportional to $r^{1/4}$, the bound \cref{eq:EI_tensor_modes} is expected to improve only modestly.
 
\paragraph{Brief review of ALP parameter space constraints.} The ALP dark matter scenario considered here has a large variety of phenomenological implications. Various lower bounds on the axion mass have been derived in the fuzzy dark matter region $m_a \sim 10^{-22} - 10^{-18}$~eV. For a recent review, see Ref.~\cite{Eberhardt:2025caq}. The generic argument, that the de Broglie wavelength of dark matter has to be smaller than the size of dwarf galaxies has recently been scrutinized~\cite{Zimmermann:2024xvd}, leading to the robust lower bound of $m_a > 2.2\times 10^{-21}$~eV. Taking into account effects on structure formation, more stringent bounds can be obtained. Ref.~\cite{Rogers:2020ltq} obtains a bound from Lyman-$\alpha$ forest constraints on the matter power spectrum of 
\begin{equation}\label{eq:ly-a_ma_bound}
m_a > 2\times 10^{-20} \,{\rm eV} \,,
\end{equation}
for previous, weaker Lyman-$\alpha$ limits, see, e.g.,~\cite{Irsic:2017yje,Kobayashi:2017jcf}. 
The limit in \cref{eq:ly-a_ma_bound} has been derived under the assumption of an initially homogeneous axion distribution, which applies to the pre-inflationary misalignment mechanism. Recently, the importance of free streaming effects due to sizeable axion momenta has been pointed out, especially in the context of post-inflationary scenarios~\cite{Amin:2022nlh,Liu:2024pjg,Ling:2024qfv,Harigaya:2025pox,Amin:2025sla,Long:2024cak}. Ref.~\cite{Liu:2024pjg} has shown that the bound from \eqref{eq:ly-a_ma_bound} can be mapped to 
\begin{equation}\label{eq:ly-a_ma_bound_post}
m_a > 5.6\times 10^{-19} \,{\rm eV}     
\end{equation}
for post-inflationary axions. In the following, we adopt the bounds \cref{eq:ly-a_ma_bound,eq:ly-a_ma_bound_post} as reference lower bound for the pre- and post-inflationary cases, respectively.
We leave a more detailed investigation of free streaming in our scenario for future work.
Even stronger limits have been derived from the stellar heating of ultra-faint dwarf galaxies~\cite{Marsh:2018zyw,Dalal:2022rmp,Teodori:2025rul,May:2025ppj}, although these limits have been criticized in \cite{Chiang:2021uvt,Yang:2025bae} due to tidal effects of the host galaxy.

Constraints of a different kind can be derived from rotating black holes (BHs) due to superradiance~\cite{Arvanitaki:2009fg,Brito:2015oca}.Light bosons extract rotational energy from spinning BHs and form bound states, so-called bosonic clouds, when their Compton wavelength matches the BH size. The absence of this effect constrains their mass. For non-interacting bosons, large ranges of $m_a$ are disfavoured (see, e.g.,~\cite{Stott:2018opm,Davoudiasl:2019nlo}) with supermassive black holes (SMBHs) covering the low-mass region roughly from $10^{-20}$ to $10^{-16}$~eV and stellar-mass black holes from around $10^{-13}$ to $10^{-11}$~eV. However, for bosons with a potential of the form in \cref{eq:V}, the resulting self-interactions, which scale as $1/f_a^2$, can prevent the formation of bosonic clouds around the black hole~\cite{Baryakhtar:2020gao,Xie:2025npy} and thereby limit the mass constraints to high values of $f_a$ \cite{Hoof:2024quk}. In our analysis, we use the excluded regions from \cite{Unal:2020jiy} for SMBHs and \cite{Witte:2024drg} for stellar-mass BHs; these constraints include the effects of self-interactions and assume a potential consistent with \cref{eq:V}. 

%%%%%%%%%%%%%%%%%%%%%%%%%%%%%%%%%%%%%%%%%%%%%%%%%
\section{Post-inflationary ALP dark matter}
\label{sec:ALP_DM_post}
%%%%%%%%%%%%%%%%%%%%%%%%%%%%%%%%%%%%%%%%%%%%%%%%%

\subsection{Axions from cosmic strings}

During phase transitions in the early Universe, topological defects can form as solitonic solutions to the equation of motion. Unlike in the pre-inflationary scenario, such defects are not diluted away by inflation in the post-inflationary Universe, allowing them to persist and potentially have cosmological consequences. When the complex scalar field spontaneously breaks the global $U(1)$ symmetry by acquiring a vacuum expectation value, $\theta$ can wind continuously from $-\pi$ to $\pi$ around certain points in space reflecting the non-trivial topology of the vacuum manifold. At the centre of such windings, the field remains trapped at the origin and the symmetry is locally restored. This leads to a configuration in which energy is concentrated along a one-dimensional defect called cosmic string with a thickness of $1/m_r$. Strings form a network whose (self-)intersections produce closed loops. These loops shrink and radiate primarily relativistic axions, which eventually get redshifted, become non-relativistic and contribute to dark matter. 

The large hierarchy between the scales of spontaneous and explicit symmetry breaking, combined with the nonlinear nature of the equation of motion, leads to the breakdown of effective field theory approaches to axion string dynamics \cite{Gorghetto2018}. As a result, the detailed evolution must be studied through numerical lattice simulations. However, such simulations also face significant challenges due to the wide separation of physical scales: the lattice spacing must be fine enough to resolve the cosmic strings, while the simulation volume must be large enough to cover multiple Hubble patches to avoid finite-volume effects. In terms of the Hubble e-folding $\log(m_r/H)$, which is commonly used in literature to characterize the time evolution of the dynamics, the most advanced simulations to date can reach values of approximately $\log(m_r/H) \simeq 10$ \cite{Benabou2025}, using techniques such as adaptive mesh refinement \cite{Buschmann2022}. For comparison, the QCD phase transition occurs at a much later time, corresponding to $\log(m_r/H) \simeq 60${--}$70$.

Despite the limited range of current simulations, their results allow for extrapolation to cosmological scales. Over time, the network is expected to converge towards an attractor solution, becoming largely insensitive to the initial conditions of the simulation~\cite{Gorghetto2018}. During the cosmic expansion, the properties of the string network are expected to follow scaling laws with a logarithmic violation of the scaling behaviour. We adopt an analytical estimate of the dark matter abundance and the isocurvature power spectrum following closely the discussion in~\cite{Gorghetto2018,Gorghetto2021,Kim2024}.

The scaling regime takes place from global $U(1)$ symmetry breaking at temperatures $T \sim f_a$ down to the temperature $T_*$, when the axion mass becomes relevant and the string network dissolves by forming domain walls, which collapse quickly. We comment on a possible additional axion contribution from this process at the end of this section. The temperature $T_*$ is defined by
\begin{equation}\label{eq:t_ast}
  m_a(T_*) = \gamma H(T_*) \,,
\end{equation}
with $\gamma$ being of order one. We have verified that values of $1\le \gamma\le 5$ lead only to small changes in our results, much smaller than other uncertainties discussed below. Therefore, we follow the convention in axion string literature and set $\gamma=1$.
We use the asterisk subscript to denote quantities evaluated at the temperature $T_*$.

During the epoch with temperatures $f_a > T > T_*$, axions are radiated from the string network. Exact scaling is defined as an evolution which maintains one string per Hubble volume on average. However, numerical simulations indicate a logarithmic violation of exact scaling \cite{Gorghetto2018,Gorghetto2021,Buschmann2022,Kim2024,Saikawa2024,Benabou2025}, and the number of strings per Hubble volume $\xi$ can be described by 
\begin{equation}\label{eq:xi}
    \xi = c_1 \log\left(\frac{m_r}{H}\right) + c_0 + \ldots\,,
\end{equation}
where $c_0$ and $c_1$ are constants characterizing the offset and slope of the scaling behaviour, respectively, and the ellipses indicate additional terms that become suppressed at late times. We adopt the simulation values from \cite{Kim2024}, setting $c_0 = -0.81$ and $c_1 = 0.21$. We have explicitly verified that varying $c_0\in[-10, 10]$ and $c_1\in[0.1, 1]$ does not affect our final conclusions on the axion mass constraints. At late times, when $\log(m_r/H) \gg 1$, the constant $c_0$ can be neglected, and the string number density significantly exceeds the value expected for exact scaling. We note that some studies do not find evidence for a logarithmic violation of exact scaling~\cite{Correia:2025nns}. While our main analysis assumes such a violation, we consider the alternative scenario in Appendix~\ref{sec:constant-string-density}.

Another key quantity is the normalized instantaneous emission function $F(k)$, which describes the momentum distribution of the radiated axions at each point in time. Simulations show that it follows a power-law, characterized by the spectral index $q$: $F(k) \propto k^{-q}$.
This holds between the IR cutoff $\sim H$ and the UV cutoff $\sim m_r/2$. The value of the spectral index $q$ is decisive in determining whether axion emission is dominated by infrared modes ($q>1$), ultraviolet modes ($q<1$), or follows the intermediate, scale-invariant case ($q=1$).

Some simulations indicate a logarithmic growth of $q$, potentially reaching values of order $\mathcal{O}(10)$ at late times \cite{Gorghetto2021, Kim2024}, while others find no evidence of a deviation from unity \cite{Buschmann2022}. The UV-dominated case $q<1$ is strongly disfavoured by current numerical simulations \cite{Gorghetto2021, Vaquero2019, Buschmann2022} and will therefore not be considered further. Since the precise evolution of $q$ remains debated, we consider both cases $q>1$ and $q=1$ in the following. 

By considering the axion emission rate as well as energy conservation in the expanding universe, the authors of \cite{Gorghetto2018} derive an approximate expression for the energy density of axions in the late-time limit $\log(m_r/H) \gg 1$. Total and differential energy density are related by
\begin{equation}
  \rho(t) = \int dk \frac{\partial\rho}{\partial k}(k, \,t) \,,
\end{equation}
and at leading order in $\log(m_r/H)$ the differential energy density for momenta in the range $k_{\mathrm{min}}$ to  $k_{\mathrm{max}}$ is given by \cite{Gorghetto2021, Kim2024}:
\begin{align}
  q>1: &\qquad
        \frac{\partial\rho}{\partial k}(t) 
        \approx \frac{8 H^2(t) \mu(t) \xi(t)}{k}  \,,
        \label{eq:drho_dk_IR} \\
  q=1: &\qquad
        \frac{\partial\rho}{\partial k}(t) 
        \approx \frac{8 H^2(t) \mu(t) \xi(t)}{k} 
        \frac{\log\frac{k}{k_{\rm min}}}{\log\frac{m_r}{H(t)}} \,. 
        \label{eq:drho_dk_q1} 
\end{align}
The infrared cutoff $k_{\mathrm{min}}$ is set by the inter-string distance, while the ultraviolet cutoff $k_{\mathrm{max}}$ is determined by the string core size:
\begin{align}\label{eq:kmin}
  k_{\rm min}(t) = x_0 \sqrt{\xi(t)} H(t) \quad \text{and} \quad  
  k_{\rm max}(t) \sim \sqrt{m_r H(t)} \,.
\end{align}
We adopt $x_0 = 10$ as our default value, motivated by the simulations from \cite{Gorghetto2021, Kim2024} and use the factor $\sqrt{\xi}$ in $k_{\rm min}$ to take into account the reduced inter-string distance for $\xi > 1$ \cite{Kim2024,Saikawa2024}. 
The exact expressions for $k_{\rm max}$ differ in refs.~\cite{Gorghetto2021, Kim2024}. However, $k_{\rm max}$ is irrelevant for our calculations, as late-time cosmological effects are dominated by the IR part of the spectrum. The string tension $\mu$, which quantifies the energy per unit length of the strings, is given by
\begin{equation}\label{eq:mu}
    \mu = \pi f_a^2 \log\left(\frac{m_r \, \eta}{H \sqrt{\xi}}\right) \,,
\end{equation}
where $\eta$ is a shape factor encoding the geometry of the string network. We adopt the canonical choice, $\eta = (4\pi)^{-1/2}$, which corresponds to a configuration of parallel strings~\cite{Gorghetto2021}. 
 
We see that for $q>1$ the spectrum is approximately scale-invariant being $\propto 1/k$, and independent of the exact value of $q$ in the approximation adopted in \cref{eq:drho_dk_IR}. For $q=1$, there is a logarithmic correction to scale-invariance and the spectrum is $\propto\log(k/k_{\rm min})/k$. Up to order one factors, the total energy density in both cases is given by
\begin{align}
  \rho \sim \pi f_a^2H^2 \left(\log\frac{m_r}{H}\right)^3 \,,
\end{align}
where we have used \cref{eq:xi,eq:mu}. 

Let us consider the system now at the time $t_*$ defined in \cref{eq:t_ast}. An important quantity, characterizing the end of the scaling regime, is
\begin{equation}
 \log_* \equiv \log\frac{m_r}{H_*} = \log\frac{f_a}{m_*} \,.   
\end{equation}
Assuming that axions radiated from strings provide all dark matter, we find typical values in the range $\log_* \simeq 110$ for $m_a \simeq 10^{-20}$~eV to $\log_* \simeq 50$ for $m_a = 10^{-2}$~eV.
First, we note that at $t_*$, all axions are still relativistic, as $k_{\rm min,*} = x_0\sqrt{\xi_*} m_* > m_*$. Second, the total energy density in axions is of order $\rho_* \sim \pi m_*^2f_a^2\log_*^3$. This can be compared to the potential energy from \cref{eq:V}, $\rho_{\rm pot} \lesssim m_*^2f_a^2$, which is bounded due to the cosine shape of the potential. We see that $\rho_* \sim \rho_{\rm pot}\log_*^3 \gg \rho_{\rm pot}$, which means that around $t_*$ the potential term plays no role in the evolution of the axion population. The energy density is instead dominated by gradient terms. Hence, we can describe the evolution for $t> t_*$ by just relativistically redshifting energy density and momenta. We use that for a radiation-dominated universe, $H \propto a^{-2}$ and obtain
\begin{equation} \label{eq:redshift}
  \begin{split}    
        &\frac{\partial\rho}{\partial k}(t) =  
        \frac{\partial\rho}{\partial k}(t_*) \frac{H^2(t)}{H_*^2}\\
        &k_{\rm min}(t) = k_{\rm min}(t_*) \sqrt{\frac{H(t)}{H_*}}
         = x_0 \sqrt{\xi_* H_* H(t)} 
  \end{split}
        \quad \text{for }
        t\ge t_*\,.  
\end{equation}
This evolution continues until the total energy becomes comparable to the potential energy and a nonlinear transition occurs at a temperature $T_\ell < T_*$~\cite{Gorghetto2021,Miyazaki:2025tvq,Narita:2025jeg}.
Due to the nonlinear nature of this process, numerical simulations are required to describe it accurately. Here, we apply the simulation-motivated analytical estimate from Ref.~\cite{Gorghetto2021}, which we briefly outline below.

The temperature $T_\ell$ is defined by requiring that the potential and the energy density in non-relativistic modes become comparable~\cite{Gorghetto2021}:
\begin{align}\label{eq:T_ell}
    \rho_{\mathrm{IR}}(T_\ell) \equiv
    \int_0^{c_m m_a(T_\ell)} \frac{\partial\rho}{\partial k}(T_\ell) 
    = c_v\,m_a^2(T_\ell)\,f_a^2 
    = \rho_{\mathrm{pot}}(T_\ell)\,.
\end{align}
Here, $c_m$ and $c_v$ are $\mathcal{O}(1)$ coefficients extracted from numerical studies. Based on \cite{Gorghetto2021}, we set $c_m = 2.08$ and $c_v = 0.13$. Larger (Smaller) values would increase (decrease) the resulting abundance, but do not affect the final mass limits. Once the system enters the nonlinear regime, axion number violating self-interactions for modes with momenta $k < m_a(T_\ell)$ convert their energy into massive non-relativistic axions. Consequently, a longer delay before the onset of the nonlinear regime implies a stronger suppression of the final number density: the later the transition occurs, the larger the axion mass has grown, and the more energy is required to produce each non-relativistic axion. For simplicity, we model this transition as instantaneous. Afterwards, the comoving number density is conserved.

With \cref{eq:drho_dk_q1,eq:drho_dk_IR}, \cref{eq:T_ell} becomes
\begin{align}
&q>1: \qquad  8H_\ell^2\xi_*\mu_* \log\frac{c_m m_\ell}{k_{\rm min,\ell}} = c_v f_a^2 m_\ell^2 \,, \label{eq:condTell1}\\
&q=1: \qquad  4H_\ell^2\xi_* \frac{\mu_*}{\log_*} 
\left(\log\frac{c_m m_\ell}{k_{\rm min,\ell}}\right)^2 = c_v f_a^2 m_\ell^2 \,. \label{eq:condTell2}
\end{align}
Using $H\propto T^2$ and $m_a(T)\propto T^{-\alpha/2}$, this can be rewritten as an implicit equation for $m_\ell$, which can be solved in terms of the Lambert $W$ function \cite{Gorghetto2021, Kim2024}. We follow this procedure in our numerical calculations. A very rough estimate can be obtained by setting $\log(c_m m_\ell/k_{\rm min,\ell})\sim 1$ and using $\mu_* \sim f_a^2 \log_*$ and $m_\ell/H_\ell = (m_\ell/m_*)^{1+4/\alpha}$, which leads to the parametric dependence
\begin{align}
  \frac{m_\ell}{m_*} \sim \left\{
  \begin{array}{l@{\qquad}l}
    \left(\xi_*\log_*\right)^{\frac{1}{2}\frac{\alpha}{4+\alpha}} & (q>1)\\
    \mathcal{O}(1)& (q=1)
  \end{array}
  \right. \,.
  \label{eq:m_ell_m_ast}
\end{align}

%%%%%%%%%%%%%%%%%%%%%%%%%%%%%%%%%%%%%%%%%%%%%%%%%%%%%%%%%%%%%%%%%%%%%%%%%%%%%%%
\subsection{Axion dark matter abundance}
\label{sec:DM}

The contribution to dark matter is determined by the final abundance of non-relativistic axions, which is set by the number density at the moment when the comoving number density becomes conserved. The relevance of the delay of the nonlinear transition discussed above depends significantly on the model parameters. We distinguish between the following cases:
\begin{itemize}
    \item $q>1, \, \alpha>0$: In this case, the delay of the nonlinear transition is important, since $T_\ell$ is significantly lower than $T_*$, but larger than $T_c$, where the axion reaches its zero-temperature mass. 
    For temperatures  $T < T_\ell$, the mass term becomes relevant and the comoving number density of axions is conserved (although the mass keeps growing). Once all momenta are redshifted away, the resulting non-relativistic axions contribute to dark matter. Hence, the present-day energy density of the axions radiated by cosmic strings during the scaling regime is given by
\begin{equation}
    \rho(T_0) = m_a n(T_\ell) \left(\frac{a_\ell}{a_0}\right)^3 \,,
    \label{eq:string_abundance}
\end{equation}
with $a_\ell$ ($a_0$) denoting the cosmic scale factor at $T_\ell$ (today). The axion number density at the nonlinear transition, $n(T_\ell)$, is dominated by the IR part\footnote{Using \cref{eq:drho_dk_q1,eq:drho_dk_IR}, one can show that the number of UV axions at $T_\ell$, i.e., with momenta larger than $c_m m_\ell$, is at most of the same order as the IR part or smaller.}. We can use \cref{eq:T_ell} to estimate $n(T_\ell) \approx \rho_{\rm IR}(T_\ell) / m_\ell$ and hence, 
\begin{equation}\label{eq:number_density_Tell}
    n(T_\ell) = c_n c_v m_\ell f_a^2 \,.
\end{equation}
The coefficient $c_n$ accounts for transient effects and higher mode contributions. We set $c_n = 1.35$, in accordance with the simulation results from \cite{Gorghetto2021}. As for $c_m$ and $c_v$, the precise value of $c_n$ does not impact the resulting mass constraints.

\item $q > 1, \, \alpha=0$: For a temperature-independent axion mass, we have $m_*=m_\ell=m_a$, and the nonlinear transition only slightly modifies the final abundance, as the primary effect comes from the increase of the axion mass during the delay \cite{GorghettoGW2021}. 
Hence, in this case, we can assume that the comoving number density is conserved already starting at $T_*$ and
\begin{equation}
    \rho(T_0) = m_a n(T_*) \left(\frac{a_*}{a_0}\right)^3
    \label{eq:string_abundance_alpha0}
\end{equation}
with
\begin{align}
    n(T_*) = \int_{k_{\rm min,*}}^{k_{\rm max,*}}
    dk \frac{1}{k} \frac{\partial\rho}{\partial k} \approx
    \frac{8 H^2_* \mu_* \xi_*} {k_{\rm min,*}}
    \sim m_* f_a^2 \sqrt{\xi_*}\log_* \,. 
    \label{eq:number_density_alpha0}
\end{align}

\item $q=1$: Similarly, in this case, \cref{eq:m_ell_m_ast} implies $T_\ell\simeq T_*$ (for any $\alpha \ge 0$) and we can also use \cref{eq:string_abundance_alpha0}.\footnote{The assumption of conserved axion number density from $t_*$ on for $q=1$ is consistent with the simulation results reported in the supplementary material of \cite{Benabou2025}.} However, the number density at $T_*$ is now given by
\begin{align}
    n(T_*) \approx
    \frac{8 H^2_* \mu_* \xi_*} {\log_* k_{\rm min,*}}
    \sim m_* f_a^2 \sqrt{\xi_*} \,,     
    \label{eq:number_density_q1}
\end{align}
a factor $\log_*$ smaller than \cref{eq:number_density_alpha0}.
\end{itemize}

Let us compare the dark matter abundance from string radiation to the generic expectation from the misalignment mechanism, where the comoving number density is conserved already for $T \lesssim T_*$ and $n_{\rm mis}(T_*) \sim m_* f_a^2$. For $q>1,\,\alpha >0$ we obtain
\begin{equation}\label{eq:rho_mis_nonlin}
\begin{split}    
  \frac{\rho_{\rm str}(T_0)}{\rho_{\rm mis}(T_0)} &\sim 
  \frac{n_{\rm str}(T_\ell)}{n_{\rm mis}(T_\ell)} \sim \frac{m_\ell}{m_*} \left(\frac{H_*}{H_\ell}\right)^{3/2} 
  \approx  \left(\frac{m_\ell}{m_*}\right)^{1+\frac{6}{\alpha}} \\[3mm]
     & \sim (\xi_*\log_*)^{\frac{1}{2} \frac{6+\alpha}{4+\alpha}}
    \sim \sqrt{\xi_*\log_*} 
\end{split}
    \qquad (q>1,\, \alpha > 0) \,,
\end{equation}
where we have used \cref{eq:m_ell_m_ast}. For the other two cases we find
\begin{equation}\label{eq:rho_mis_a0_q1}
  \frac{\rho_{\rm str}(T_0)}{\rho_{\rm mis}(T_0)} \sim 
  \frac{n_{\rm str}(T_*)}{n_{\rm mis}(T_*)} \sim 
  \left\{
  \begin{array}{l@{\qquad}l}
     \sqrt{\xi_*}\log_* &     (q>1,\, \alpha = 0) \,, \\[3mm]
     \sqrt{\xi_*}  &     (q=1,\, \alpha \ge 0) \,.
  \end{array}
  \right.
\end{equation}
Therefore, we expect that, independent of $\alpha$, the dark matter abundance from string radiation largely dominates over the misalignment abundance for $q>1$, whereas for $q=1$, we expect a modest enhancement. We have confirmed this by implementing the semi-analytical estimate for the misalignment mechanism from \cite{Enander:2017ogx}. We find that in the entire considered parameter space $\rho_{\rm mis} \lesssim 0.01\rho_{\rm str}$ for $q>1$, whereas for $q=1$, $\rho_{\rm mis}$ can reach up to $ \sim 0.25\rho_{\rm str}$. In the following, we neglect the misalignment mechanism and assume that string radiation is the dominant dark matter axion production mechanism.

\begin{figure}[!t]
\centering
    \includegraphics[width=0.8\textwidth]{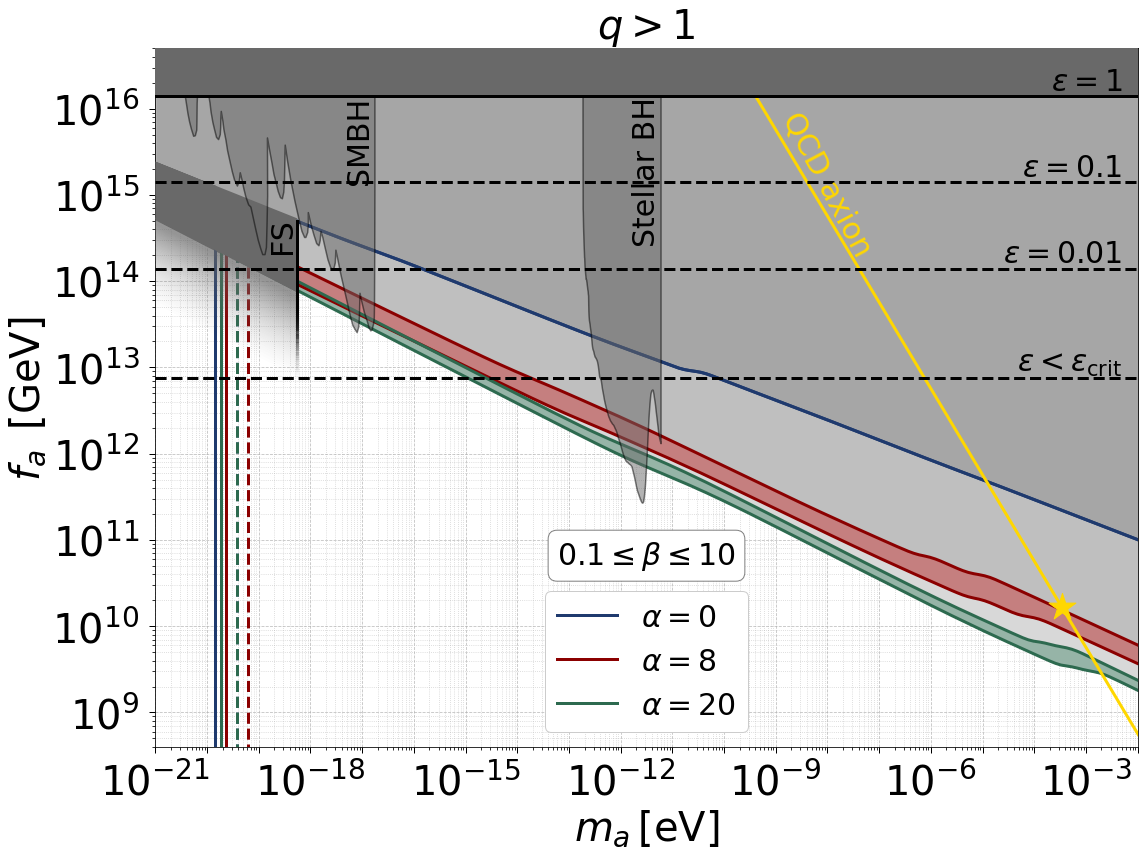} 
    \caption{Axion decay constant $f_a$ vs. axion mass $m_a$ in the post-inflationary scenario, for different $\alpha,\,\beta$ and for $q > 1$, assuming that axions radiated by cosmic strings during the scaling regime constitute all of dark matter. Shaded regions above the curves overproduce dark matter. The band width reflects $\beta$ = 0.1--10, with the lower (upper) boundary corresponding to 0.1 (10). The IR momentum cutoff is $k_{\rm min} = x_0\sqrt{\xi_*H_*H}$ with $x_0 = 10$. Vertical solid (dashed) lines indicate the lower bound on $m_a$ from isocurvature constraints using UFDs (21~cm), see \cref{sec:iso-post}. The blue and green dashed lines overlap. Horizontal lines indicate the upper bound on $f_a$ from CMB tensor modes, depending on the reheating efficiency $\epsilon$ with $\epsilon_{\rm crit} = 5.4\times 10^{-4}$, see \cref{sec:tensor-modes}. Also shown are the lower bound on $m_a$ from free streaming (FS), \cref{eq:ly-a_ma_bound_post} \cite{Liu:2024pjg}, regions excluded by BH superradiance \cite{Unal:2020jiy,Witte:2024drg}, and the QCD axion (yellow line) with the star marking where the QCD-like $m_a(T)$ gives the correct dark matter abundance.}
    \label{fig:strings_abundance_q3}
\end{figure}

\begin{figure}[!t]
\centering
    \includegraphics[width=0.8\textwidth]{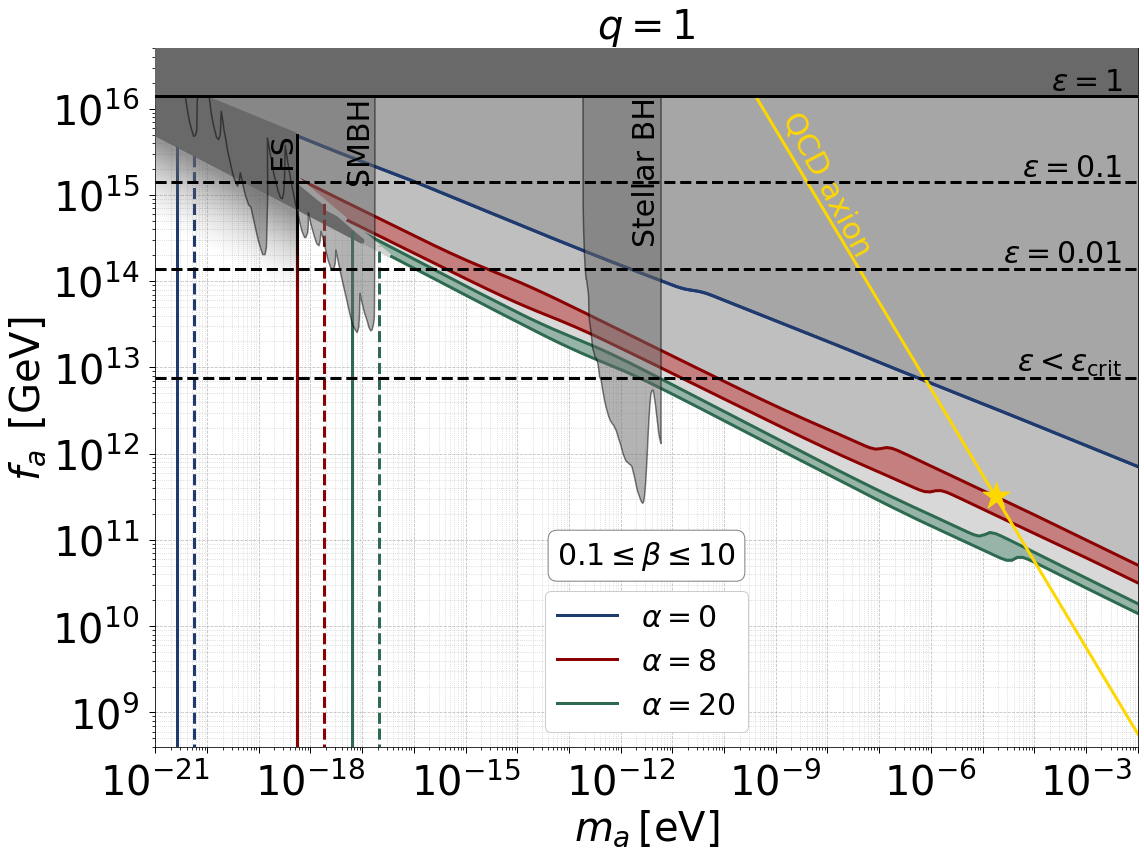}
    \caption{Same as \cref{fig:strings_abundance_q3} but for a scale-free axion emission spectrum ($q=1$).}
    \label{fig:strings_abundance_q1}
\end{figure}

In \cref{fig:strings_abundance_q1,fig:strings_abundance_q3}, we present some results of our numerical analysis. We show, as a function of the zero-temperature axion mass $m_a$, the value of $f_a$ required for the axion abundance produced by cosmic strings during the scaling regime to account for all of dark matter. The figures show the dependence of the results on the mass-temperature parametrization parameters $\alpha$, $\beta$ and the spectral index $q$ of the axion emission spectrum, with $q>1$ illustrated in \cref{fig:strings_abundance_q3}
and $q=1$ in \cref{fig:strings_abundance_q1}. The grey-shaded region above a given curve leads to an overabundance of dark matter. The reason for the kink in the curves, also visible in many of the following graphs, is the proximity of $T_*$ to the QCD phase transition temperature and the related sudden drop in relativistic degrees of freedom.
Several trends are evident. The parameter $\beta$ has a minor impact compared to $\alpha$. Smaller values of $\alpha$ correspond to higher $T_*$, and therefore a smaller redshift factor $(a_*/a_0)^3$, which needs to be compensated by a larger $f_a$ to keep the dark matter abundance constant. Note that \cref{eq:rho_mis_nonlin,eq:rho_mis_a0_q1} suggest an enhanced dark matter abundance for $\alpha=0$ compared to $\alpha > 0$ (for $q>1$). However, the redshift factor related to a higher $T_*$ and therefore earlier time for $\alpha=0$ is more important than the $\sqrt{\log_*}$ enhancement, explaining larger values of $f_a$ for $\alpha=0$ than for $\alpha>0$. 

For $q = 1$ (\cref{fig:strings_abundance_q1}), higher values of $f_a$ are required than for $q > 1$ (\cref{fig:strings_abundance_q3}) due to the $\sqrt{\log_*}$ ($\log_*$) enhancement for $\alpha > 0$ ($\alpha = 0$) in the latter case, see \cref{eq:rho_mis_nonlin,eq:rho_mis_a0_q1}. The precise value of $q$ has little effect as long as $q > 1$. We have verified that variations in $k_\mathrm{min}$ have a negligible effect when $q > 1$ and $\alpha = 8$ \cite{Kim2024}, while moving away from this benchmark model, we observe that changing $k_{\rm min}$ can have a notable impact on the predicted axion relic abundance, as suggested by the estimates in \cref{eq:number_density_Tell,eq:number_density_alpha0,eq:number_density_q1}. For instance, for $\alpha=0$ or $q=1$, dropping the factor $\sqrt{\xi_*}$ in $k_{\rm min}$ changes the value of $f_a$ required to obtain the dark matter abundance for fixed $m_a$ by a factor $2-3$.

\bigskip

Let us add the following comments. In our calculations, the spectra in \cref{eq:drho_dk_IR,eq:drho_dk_q1} are cut off and set to zero for $k < k_{\rm min}$. However, simulations  show a power-law behaviour for $k < k_\mathrm{min}$, approximately proportional to $k^3$ \cite{Gorghetto2021}. We have checked explicitly that including such an IR tail does neither affect our results for the axion dark matter abundance nor the isocurvature power spectrum discussed below. Therefore, we neglect this part of the energy spectrum.

The semi-analytical approach described above has been developed in refs.~\cite{Gorghetto2018,Gorghetto2021,Kim2024} for the QCD axion case. Here, we apply the same calculation to general axion-like particles, considering $f_a$ and $m_a$ to be independent parameters as well as allowing for a general temperature dependence of the axion mass. In particular, we adopt the same $\mathcal{O}(1)$ fudge factors fitted to simulations. Some discussion about the applicability of this approach can be found in \cite{GorghettoGW2021}. While a detailed investigation of this issue is beyond the scope of this work, we do study the effect of some numerical quantities extracted from simulations, for instance the precise definition of the cutoff $k_{\rm min}$. Nevertheless, one should keep in mind that this introduces an uncertainty in our results.

In addition to the axions produced during the scaling regime, strings will continue to radiate axions after $T_*$. Once the axion acquires a mass, cosmic strings become the boundaries of domain walls that interpolate between the discrete vacua of the axion potential.  A key cosmological parameter is the \textit{domain wall number} $N_{\mathrm{DW}}$, which denotes the number of degenerate minima in the potential or, equivalently, how many domain walls attach to each string. For $N_{\mathrm{DW}} \neq 1$, the resulting domain walls are stable and would rapidly dominate the energy density of the Universe, leading to the well-known \textit{domain wall problem}. This cosmological catastrophe can be avoided by introducing small explicit symmetry-breaking terms or by ensuring that the underlying model satisfies $N_{\mathrm{DW}}=1$. In the present work, we adopt the latter assumption. At the temperature $T_\ell$, the thickness of domain walls, roughly given by $m_a^{-1}$, becomes smaller than their typical separation, causing the string–wall network to decay, releasing additional axions, see e.g.,~\cite{Gorghetto2021,Benabou2025,OHare:2021zrq}. The properties of axions radiated after the scaling regime and during the collapse of the string–wall network remain highly uncertain. We therefore do not include these contributions in our analysis. Assuming they do not destructively interfere with the previously produced population, neglecting them and only focusing on axions emitted until $T_\ast$ and their subsequent evolution provides a lower bound on the total axion abundance. 

Finally, let us note that in general, any sizeable additional axion dark matter production mechanism beyond the one considered here would lower the required value of $f_a$, and therefore could potentially relax some of the constraints discussed in the following.

%%%%%%%%%%%%%%%%%%%%%%%%%%%%%%%%%%%%%%%%%%%%%%%%%%%%%%%%%%%%%%%%%%%%%%%%%%%%%%%%%%
\section{Isocurvature constraints for post-inflationary ALP dark matter}
\label{sec:iso-post}
%%%%%%%%%%%%%%%%%%%%%%%%%%%%%%%%%%%%%%%%%%%%%%%%%%%%%%%%%%%%%%%%%%%%%%%%%%%%%%%%%%

\subsection{Isocurvature power spectrum}
\label{sec:P_iso-post}

The dark matter production mechanism in the post-inflationary scenario intrinsically leads to spatial variations in the axion number density, giving rise to isocurvature fluctuations \cite{Feix2019}. The typical size of the fluctuations is determined by the dynamics around the time when axions start evolving as dark matter. Due to causality, the density fluctuations are expected to follow a flat (white-noise) power spectrum at low momenta up to a characteristic cutoff scale.

The power spectrum of dark matter density fluctuations $P_\delta(k)$ quantifies the relative contribution of different momentum modes to the total variance of the density field. Although related, $P_\delta(k)$ contains more information than the simple density spectrum $\partial \rho/\partial k$, and to relate these two quantities an additional assumption is needed on the correlation between momentum modes. Requiring statistical homogeneity and isotropy, we assume that modes with different three-momenta $\vec k$ are uncorrelated.\footnote{In Ref.~\cite{Vaquero2019}, the power spectrum following from this assumption has been compared to a direct numerical simulation and an agreement up to $\mathcal{O}(1)$ factors has been found. A possible origin of these deviations may be the presence of structures of size comparable to the simulation box, which corresponds to violations of statistical homogeneity and isotropy.} This assumption is implemented by discretizing the momentum modes and assuming a random phase for each mode. We provide details of the calculation in \cref{sec:random-phase-model}. Here, we only state the result for the power spectrum in terms of the spectral density:
\begin{equation}\label{eq:P_iso}
    P_{\delta}(k) = \frac{\pi^{2}}{\bar\rho^2_{{\rm DM,}a}} \, \frac{1}{k} \int_{k_{\rm min}}^{k_{\rm max}} dl \frac{1}{l} \frac{\partial \rho_{{\rm DM},a}}{\partial l} \int_{\lvert l - k\rvert}^{\lvert l + k\rvert} dl^{\prime} 
    \frac{1}{l^{\prime}} \frac{\partial \rho_{{\rm DM},a}}{\partial l^{\prime}} \,,
\end{equation}
with the average axion dark matter density
\begin{equation}\label{eq:rho-bar}
  \bar\rho_{{\rm DM},a} =  \int_{k_{\rm min}}^{k_{\rm max}} dk \frac{\partial \rho_{{\rm DM},a}}{\partial k} \,.  
\end{equation}
Expanding for small momenta $k$, we obtain
\begin{equation}\label{eq:P0}
P_0 \equiv  P_{\delta}(k\to 0) = \frac{2\pi^{2}}{\bar\rho^2_{{\rm DM,}a}}  \int_{k_{\rm min}}^{k_{\rm max}} dl
  \left(\frac{1}{l} \frac{\partial \rho_{{\rm DM},a}}{\partial l} \right)^2 \,,
\end{equation}
which is independent of $k$ and hence describes a white-noise power spectrum. The small-$k$ expansion holds for $k \ll l$, so that we expect the cutoff for $k \gtrsim k_{\rm min}$. 
Here, $\partial \rho_{{\rm DM},a}/\partial k$ is the axion dark matter spectrum at late times, when all axions are non-relativistic. Hence, it is proportional to the axion number density spectrum at the time when momentum modes start to evolve freely. The shape of the spectrum depends again on the spectral index $q$ and the axion mass temperature dependence $\alpha$. 

\begin{figure}[!t]
    \centering
    \includegraphics[width=0.6\textwidth]{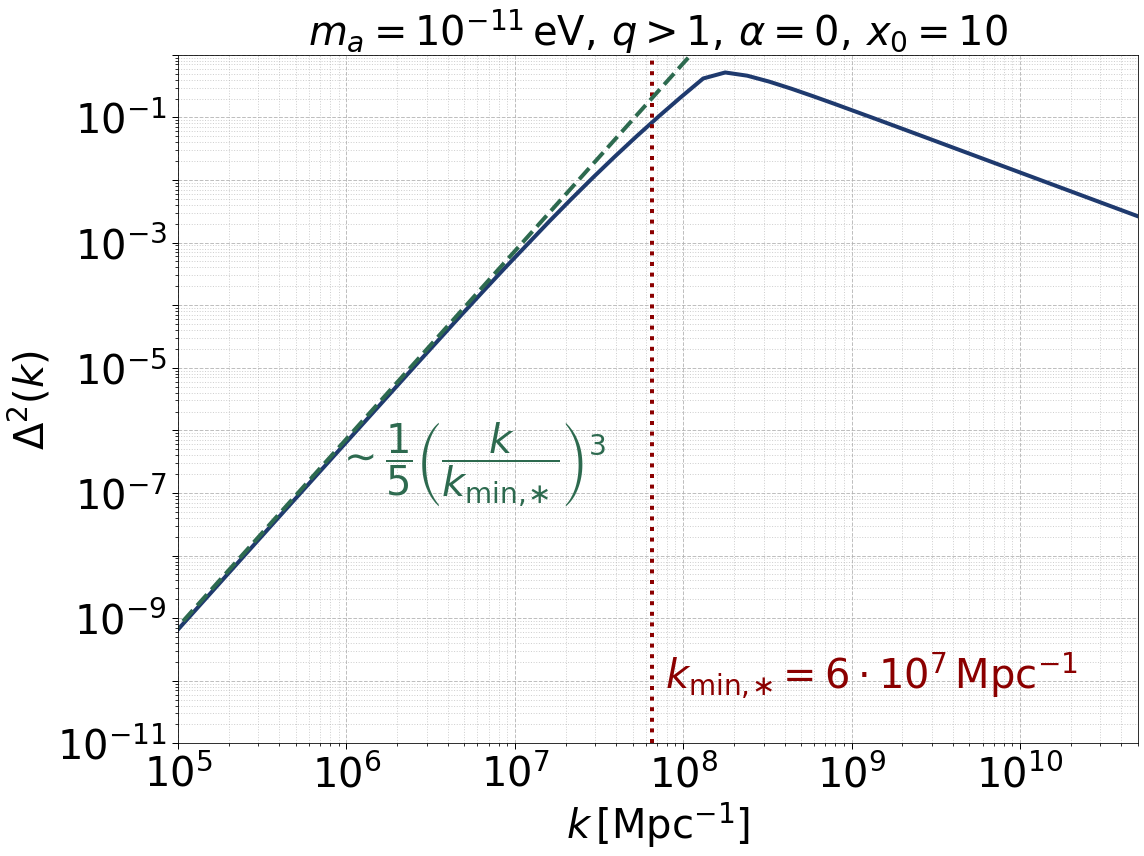}
    \caption{Isocurvature power spectrum according to \cref{eq:P_iso} for $m_a = 10^{-11}$~eV and model parameters as indicated in the figure. 
    $f_a \simeq 10^{13}$~GeV has been determined from requiring full dark matter abundance. The dotted line indicates the corresponding value of the IR cutoff $k_{\rm min,\ast}$. The dashed line corresponds to the analytic expression for small $k$ from \cref{eq:Delta}. }
    \label{fig:Delta_kmin}
\end{figure}

\begin{figure}[!t]
    \includegraphics[width=0.48\textwidth]{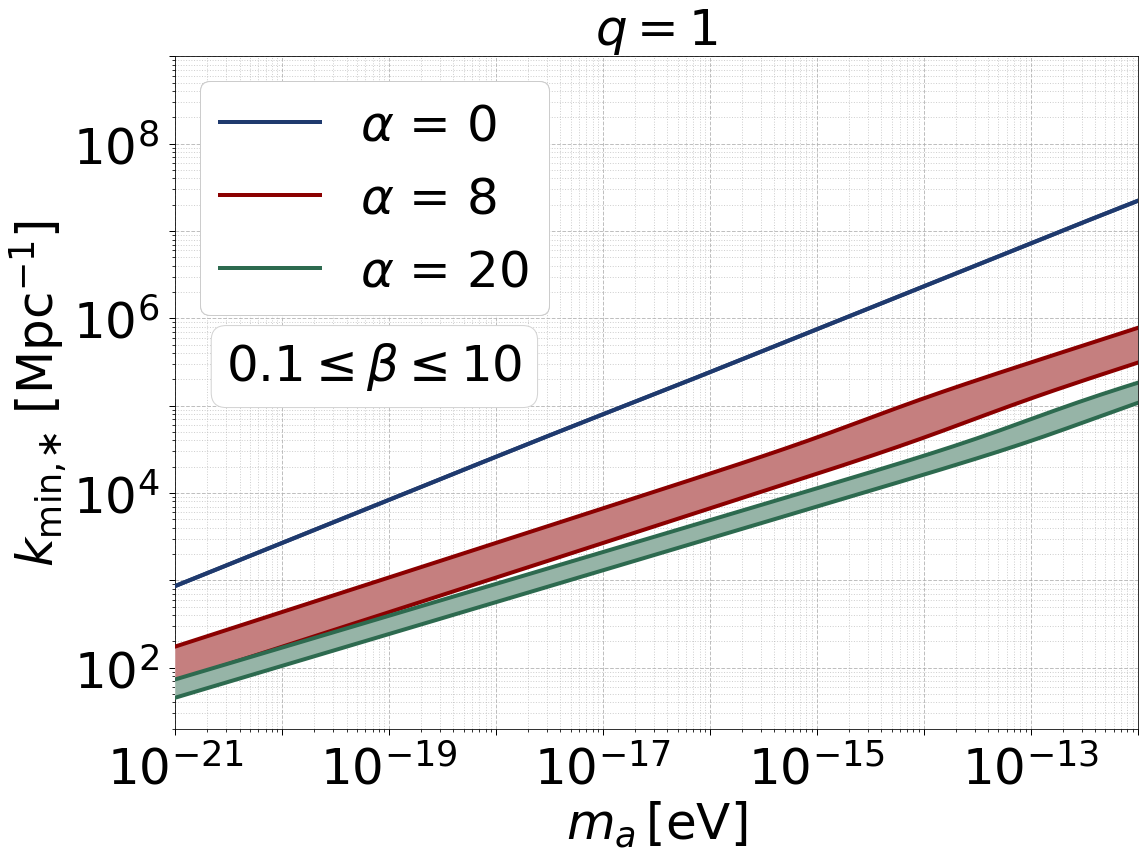}\quad
    \includegraphics[width=0.48\textwidth]{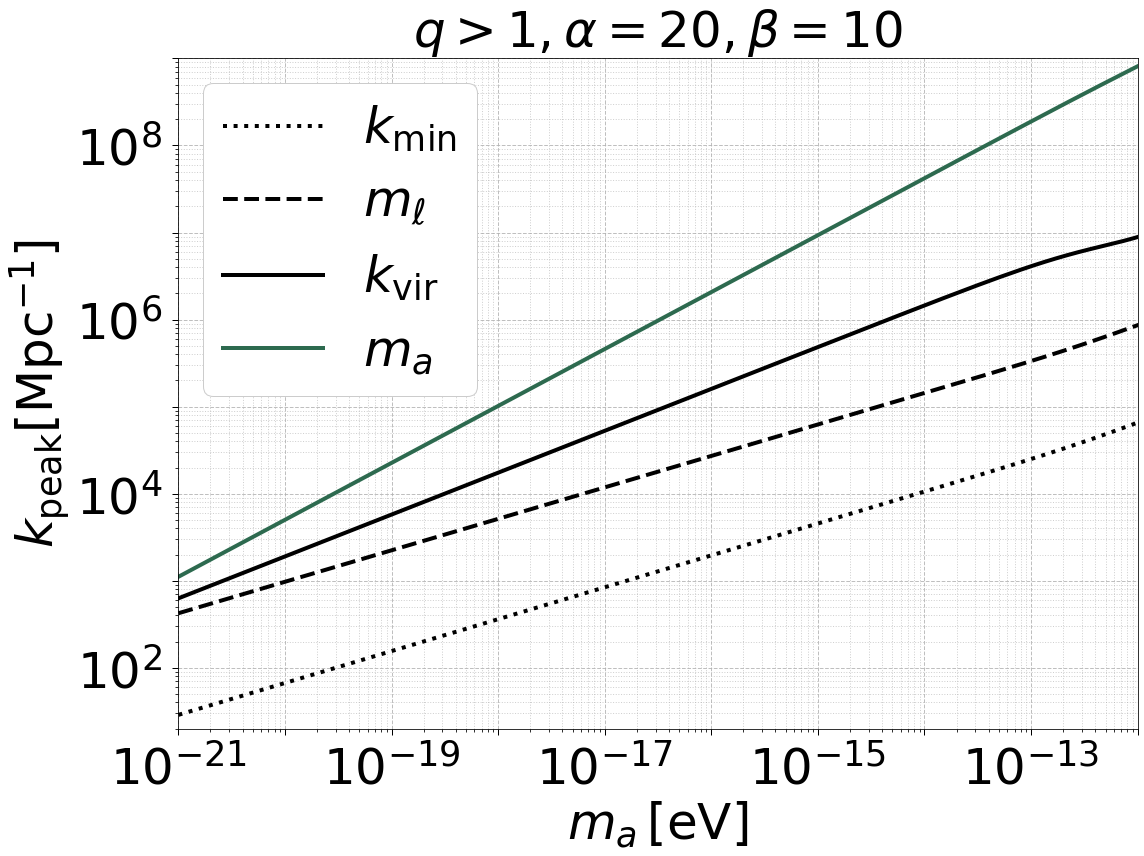}
    \caption{Comoving size of the IR spectrum cutoff as a function of $m_a$, with $f_a$ determined by requiring full dark matter abundance.
    \textit{Left:} $k_{\mathrm{min},*}$ for $q=1$ and various choices of $\alpha$ and $\beta$. The parameter $x_0$ is fixed to 10; modifying its value leads to an approximately linear rescaling of $k_{\mathrm{min},*}$. \textit{Right:} Comparison of $k_{\mathrm{min},*}$ to the peak momentum $k_{\rm peak}= m_\ell$ after the nonlinear transition at $T_\ell$ and $k_{\rm peak}= k_{\rm vir}$ after the nonlinear transition due to self-interactions at $T_c$. For comparison, we also show the comoving momentum corresponding to $m_a$ at $T_c$. 
    \label{fig:IRcutoff}}
\end{figure}

For $q=1$ or $\alpha=0$, the spectrum remains unperturbed for $t \gtrsim t_*$ apart from redshifting. Using that axions are relativistic at $T_*$, we have
\begin{equation}\label{eq:rho_DM_q1}
    \frac{\partial \rho_{{\rm DM},a}}{\partial k}(T_*) = 
    \frac{m_a}{k} \, \frac{\partial\rho} {\partial k} (T_*) \qquad\qquad
          (q=1 \quad\text{or}\quad \alpha=0) \,,
\end{equation}
and $\partial\rho / \partial k$ is given by \cref{eq:drho_dk_IR,eq:drho_dk_q1} evaluated at $t_*$. In these cases the cutoff is set by $k_{\rm min,*}$, see \cref{eq:kmin}. The integrals in \cref{eq:rho-bar,eq:P0} can be performed and we find
\begin{equation}\label{eq:P0ana}
  P_0 =  c_P \frac{2\pi^{2}}{k_{\rm min,*}^3} \,, \qquad
   c_P = \left\{ \begin{array}{l@{\quad}l}
    1/5 &  (q>1,\,\alpha = 0)\\
    2/125 & (q=1)
\end{array} \right. \,.
\end{equation}  
The dimensionless power spectrum is given by
\begin{equation}\label{eq:Delta}
    \Delta^2_a(k) \equiv \frac{k^3}{2\pi^2}P_\delta(k) \approx
    c_P \left(\frac{k}{k_{\rm min,*}}\right)^3 \qquad
    (k \ll k_{\rm min,*}) \,.
\end{equation}
Note that in \cref{eq:P0ana,eq:Delta}, $k_{\rm min}$ is evaluated at $t_*$ and then redshifted to its comoving (present-time) size. We see that fluctuations scale with $k^3$, become of order one for $k\sim k_{\rm min}$ and are suppressed again for $k \gg k_{\rm min}$. This behaviour is confirmed in  \cref{fig:Delta_kmin}, where we show an example for a power spectrum calculated from \cref{eq:P_iso}. The left panel of \cref{fig:IRcutoff} shows the relevant values of $k_{\rm min},*$ for $q=1$ and a few choices for $\alpha$ and $\beta$. 

If $q>1$ and $\alpha>0$, nonlinear transitions become important and modify the spectrum after $t_*$. First, the nonlinear transition at $t_\ell$ described in the previous section, shifts the peak of the spectrum from $k_{\rm min}$ to $k_{\rm peak}\simeq m_\ell$, see \cite{Gorghetto2021}. Second, ref.~\cite{Gorghetto:2024vnp} showed through dedicated lattice simulations of the QCD axion, starting from the time when the axion becomes non-relativistic, that the increasing axion mass for $t>t_\ell$ causes the potential energy due to axion self-interactions to dominate over the kinetic energy. This leads to the virialisation of axion overdensities at a scale set by the self-interaction potential. As shown in \cref{app:self-int}, this happens if $\alpha > 2$. While this process does not change the comoving number density of axions, the simulations of ref.~\cite{Gorghetto:2024vnp} showed that it shifts the peak of the spectrum further to higher momenta. This second nonlinear transition ends, when the axion reaches its zero-temperature mass, at $T_c = \beta^{2/\alpha}\sqrt{m_a f_a}$. At that point, the axion spectrum will peak around $k_{\rm peak}\simeq k_\mathrm{vir}$, set by the axion self-interaction coupling $\lambda_a = m_a^2/f_a^2$~\cite{Gorghetto:2024vnp}:
\begin{equation}\label{eq:k_peak}
    k_{\rm vir} \simeq \sqrt{\lambda_a a^2(T_c)} \simeq 
    %\sqrt{c_n c_v} \, 
    m_a\left(\frac{T_c}{T_\ell}\right)^{\frac{6+\alpha}{4}} \,,
\end{equation}
where $a(T_c)$ is the axion field value at the temperature $T_c$.\footnote{The non-relativistic axion dynamics are numerically much simpler to evolve than the axion string network. Therefore, the relevant lattice simulations~\cite{Gorghetto:2024vnp} do not require extrapolation over many orders of magnitude.}
We provide a more detailed discussion of this second nonlinear transition in \cref{app:self-int}. In the right panel of \cref{fig:IRcutoff}, we show the impact of the two nonlinear transitions on the peak of the spectrum. We see that $k_{\rm vir}$ from \cref{eq:k_peak} is significantly higher than the IR cutoff without nonlinear transitions $k_{\rm min}$.

Self-interactions may change the form of the spectrum around the peak \cite{Gorghetto:2024vnp}, but the value of $P_0$, i.e., the normalization of the power spectrum for small $k$, is rather insensitive to the precise form and is essentially dominated by the value of $k_{\rm peak}$. We approximate the energy spectrum at $T_c$ by a power law with an IR cutoff at $k_{\rm peak}$: 
\begin{equation}\label{eq:rho_DM_2nd-NL}
    \frac{\partial \rho_{{\rm DM},a}}{\partial k}(T_c) \propto 
    \frac{m_a}{\sqrt{m_a^2 + k^2}} \, \frac{1} {k^r} 
         \qquad  (q>1,\,\alpha>0) \,,
\end{equation}
with $k>k_{\rm peak}$, $r\gtrsim 1$. The square-root factor takes into account that the late-time dark matter density is proportional to the axion number density.  
Using this expression in \cref{eq:P0}, we find
$P_0 =c_P 2\pi^2/k_{\rm peak}^3$ with $0.02 \lesssim c_P \lesssim 0.2$. For the sake of definiteness, we choose $r = 1.2$. We have checked that $c_P$ varies by at most a factor of $\simeq 3$ for $r\in[1.0,\,1.4]$, which does not affect our final mass constraints. 

%%%%%%%%%%%%%%%%%%%%%%%%%%%%%%%%%%%%%%%%%%%%%%%%%%%%%%%%%%%%%%%%%%%%%%%%%%%%%%%%%%%
\subsection{Observational constraints and limits on the ALP parameter space}

We can now compare the predictions from the previous subsection with the observationally allowed size of isocurvature perturbations. The conventional quantity to consider is the square-root of the ratio of the power spectra of isocurvature and adiabatic perturbations at the CMB pivot scale $k_\mathrm{CMB} = 0.05 \,\mathrm{Mpc^{-1}}$:
\begin{equation}
f_{\mathrm{iso}} \equiv \left. \frac{\Delta_a}{\Delta_\mathcal{R}} \right|_{k = k_{\mathrm{CMB}}} \,.
\end{equation}
The dimensionless power spectrum of adiabatic perturbations is given by 
\begin{equation}
\Delta_\mathcal{R}^2(k) = A_s \left( \frac{k}{k_{\mathrm{CMB}}}\right)^{n_s-1}
\end{equation}
with $A_s = 2.2\times 10^{-9}$ \cite{Planck2020Parameters} and $n_s \simeq 1$. As we evaluate 
$\Delta_\mathcal{R}$ only at $k = k_{\rm CMB}$, the precise value of $n_s$ does not enter. We obtain
\begin{equation}\label{eq:fiso_pred}
f_{\mathrm{iso}}  = \sqrt{\frac{P_0 k_{\mathrm{CMB}}^{3}}{2\pi^2A_s}}  
\end{equation}
with $P_0$ given by \cref{eq:P0}. \Cref{eq:fiso_pred} holds for $k_{\rm CMB} < k_{\rm min}$.

Various cosmological constraints on $f_{\rm iso}$ in the post-inflationary axion dark matter scenario have been discussed in, e.g., \cite{Feix2019,Irsic2020,Feix2020,Shimabukuro:2020tbs,Kadota:2020ybe,Gorghetto:2025uls}.
For instance, Ref.~\cite{Feix2020} obtained a (relatively weak) limit of $f_\mathrm{iso} \leq 0.64$ at $2\sigma$ from Planck CMB data~\cite{Planck2020Overview, Planck2020Parameters}. Future CMB surveys such as CMB-S4 \cite{Abazajian2022} combined with HI 21 cm intensity mapping are expected to tighten these bounds to below 0.02~\cite{Feix2019}. While the CMB constrains modes around the pivot scale $k_\mathrm{CMB} = 0.05$~Mpc$^{-1}$, much stronger bounds can be obtained from observables that probe smaller scales due to the $\propto k^3$ dependence up to the cutoff.\footnote{For a discussion about the problem of nonlinearities, we refer to~\cite{Feix2020}.} For example, the authors of Ref.~\cite{Murgia:2019duy} derive a strong bound from Lyman-$\alpha$ forest observations in the context of primordial black holes: $f_\mathrm{iso} \leq 0.004$ at $2\sigma$, based on wavenumbers $k$ roughly between 0.1 and 10~Mpc$^{-1}$. This bound has been applied to post-inflationary ALPs in \cite{Irsic2020}.  Recently, the authors of Ref.~\cite{Graham:2024hah} considered the dynamical heating of stars in ultra-faint dwarf galaxies due to the gravitational scattering with dark matter clumps, as expected in models with large isocurvature fluctuations at small scales. For wavenumbers $k$ between 10 and 1000~Mpc$^{-1}$, they set a limit of $f_{\rm iso} \lesssim 6\times 10^{-4}$. 

Future probes include HI 21~cm intensity mapping, cosmic shear, galaxy clustering, and CMB lensing \cite{Feix2019,Irsic2020,Feix2020,Shimabukuro:2020tbs,Kadota:2020ybe}. In the following, we will adopt as our reference benchmark an optimistic estimate for the sensitivity of future 21~cm measurements from Ref.~\cite{Irsic2020}, based on the shock heating of the baryonic gas due to supersonic dark matter halos. This effect is dominated by halos with masses $M \sim 10^6$ solar masses, which corresponds to comoving wavenumbers of $k \sim (4\pi/3 \, \rho_{\rm DM,tot} /M)^{1/3} \sim 50$~Mpc$^{-1}$. 
We summarize the constraints we use in \cref{tab:fiso-limits}. 

\begin{table}[!t]
    \centering
    \begin{tabular}{cccc}
    \hline\hline
       $f_{\rm iso}$ upper limit  &  $k$ range [Mpc$^{-1}$] & label & reference \\
    \hline   
        $4\times 10^{-3}$ & $\lesssim 10$ & Lyman-$\alpha$ & \cite{Murgia:2019duy,Irsic2020} \\
        $6\times 10^{-4}$ & $10 - 1000$ & UFD & \cite{Graham:2024hah} \\
        $3\times 10^{-4}$ & $\sim 50$ & 21 cm & \cite{Irsic2020} \\
    \hline\hline    
    \end{tabular}
    \caption{Constraints on $f_{\rm iso}$ adopted in our analysis and the relevant range of wavenumbers~$k$. Lyman-$\alpha$ and ultra-faint dwarf (UFD) galaxy constraints are actual upper bounds, whereas ``21~cm'' is a sensitivity forecast based on the shock heating of baryonic gas due to dark matter halos with masses $M \sim 10^6\,M_\odot$.}
    \label{tab:fiso-limits}
\end{table}

\begin{figure}[!t]
    \includegraphics[width=0.48\textwidth]{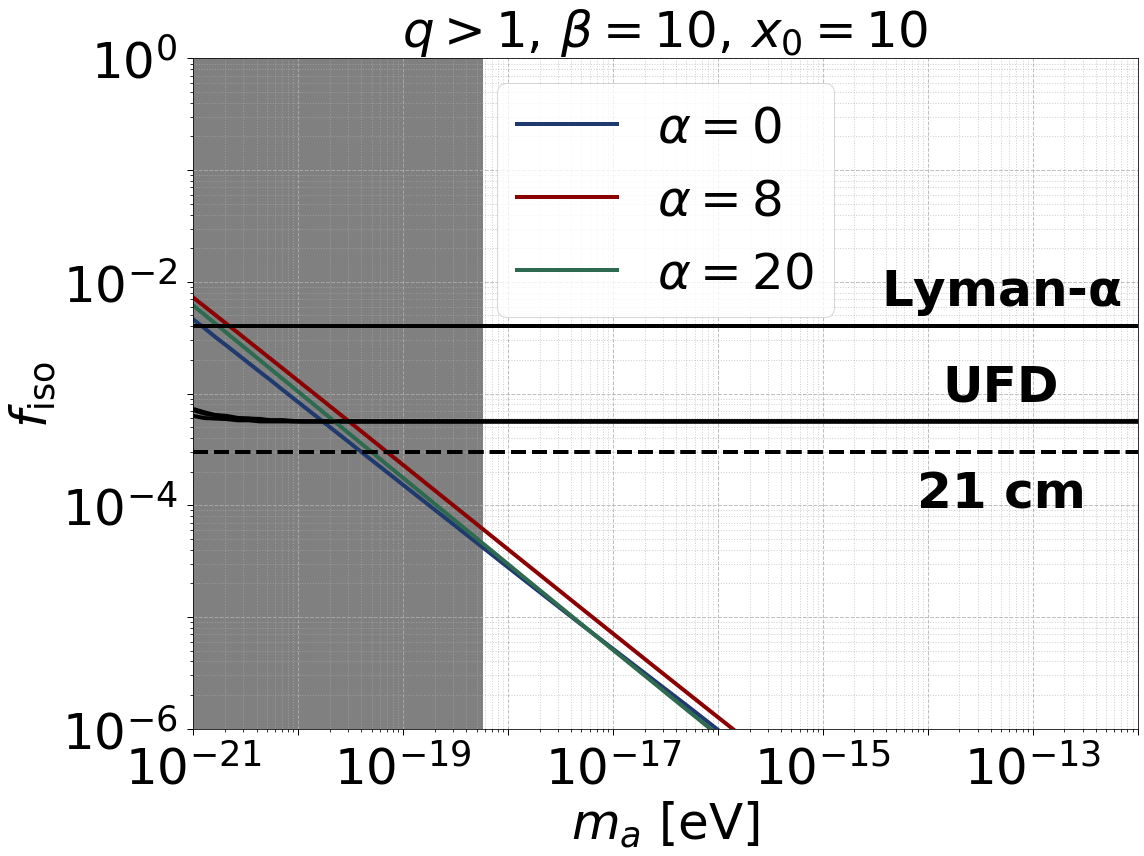}\quad
    \includegraphics[width=0.48\textwidth]{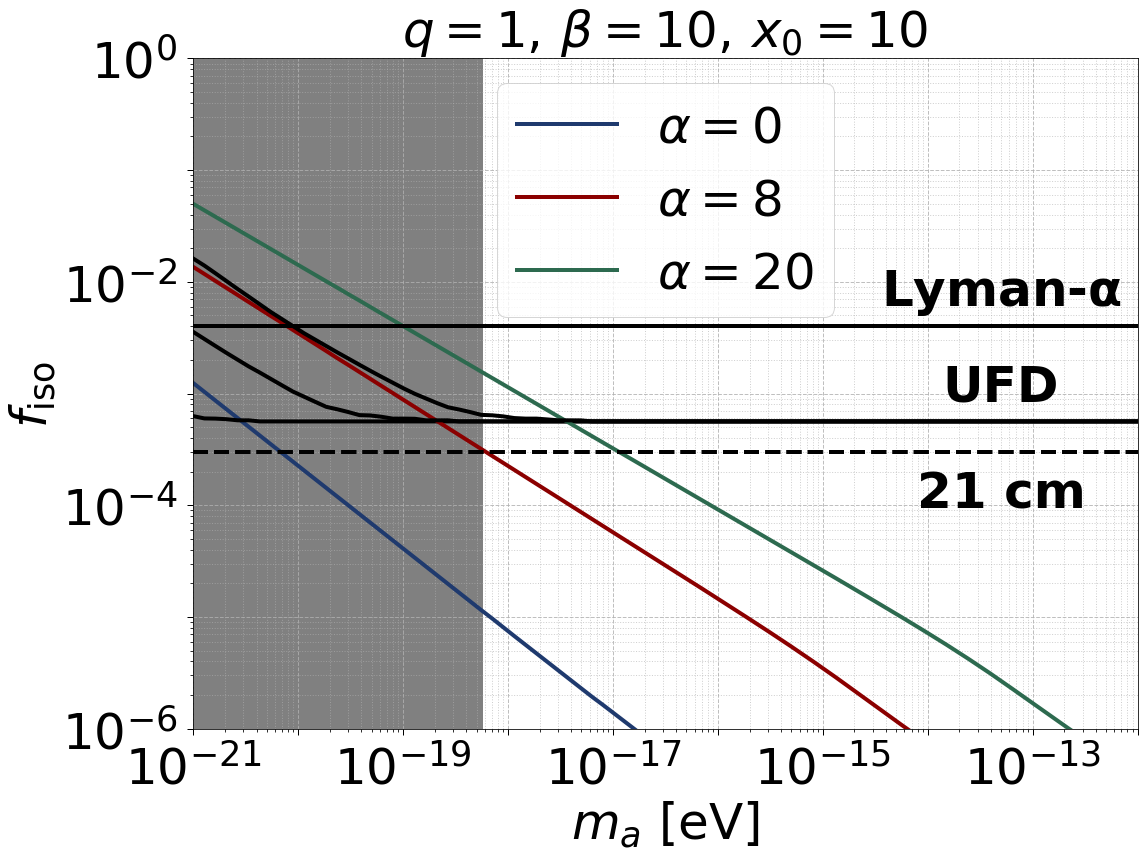} 
    \caption{Predicted isocurvature fraction $f_{\rm iso}$ vs. axion mass $m_a$, assuming 100\% dark matter abundance for different parameter choices $q,\,\alpha\text{ and } \beta$, compared to the upper bounds from Lyman-$\alpha$ and UFD (black solid), as well as 21~cm sensitivity (black dashed), see \cref{tab:fiso-limits}. The grey-shaded region indicates the lower bound on $m_a$ from free streaming, \cref{eq:ly-a_ma_bound_post}~\cite{Liu:2024pjg}. For low $m_a$, the UFD constraint weakens, depending on $q,\,\alpha,\,\beta$, but in all cases it intersects the corresponding prediction roughly at the same value of $f_{\rm iso} = 6\times 10^{-4}$.}
    \label{fig:fiso}
\end{figure}

In \cref{fig:fiso}, we show the predicted value of $f_{\rm iso}$ as a function of $m_a$ compared to these constraints for several parameter choices. The constraints apply, if the relevant $k$-range (see \cref{tab:fiso-limits}) is below the peak of the power spectrum. Hence, for each axion mass, we verify whether this condition is satisfied. We see from \cref{fig:IRcutoff}, that in the relevant regions of the parameter space, $k_{\rm peak} \gtrsim 50$~Mpc$^{-1}$, and therefore, Lyman-$\alpha$ and 21~cm constraints apply. 
%Only when $x_0$ is decreased from its default value of 10, $k_{\rm min}$ becomes too small for low $m_a$, leading to the cutoffs of the limits seen in the bottom-right panel of \cref{fig:fiso}. 
For the UFD limit, we check for each parameter choice whether the predicted isocurvature power spectrum falls within the excluded region from figure~7 of \cite{Graham:2024hah}. This leads to the weakening of the UFD constraints visible for low $m_a$. However, the predicted values of $f_{\rm iso}$
for each $(q,\,\alpha,\,\beta)$-combination shown in the plot intersects the corresponding UFD bound for $m_a$ values above the weakening of the bound, roughly at the same value of $f_{\rm iso} = 6\times 10^{-4}$.

\enlargethispage{\baselineskip}
\begin{figure}[!t]
    \includegraphics[width=0.48\textwidth]{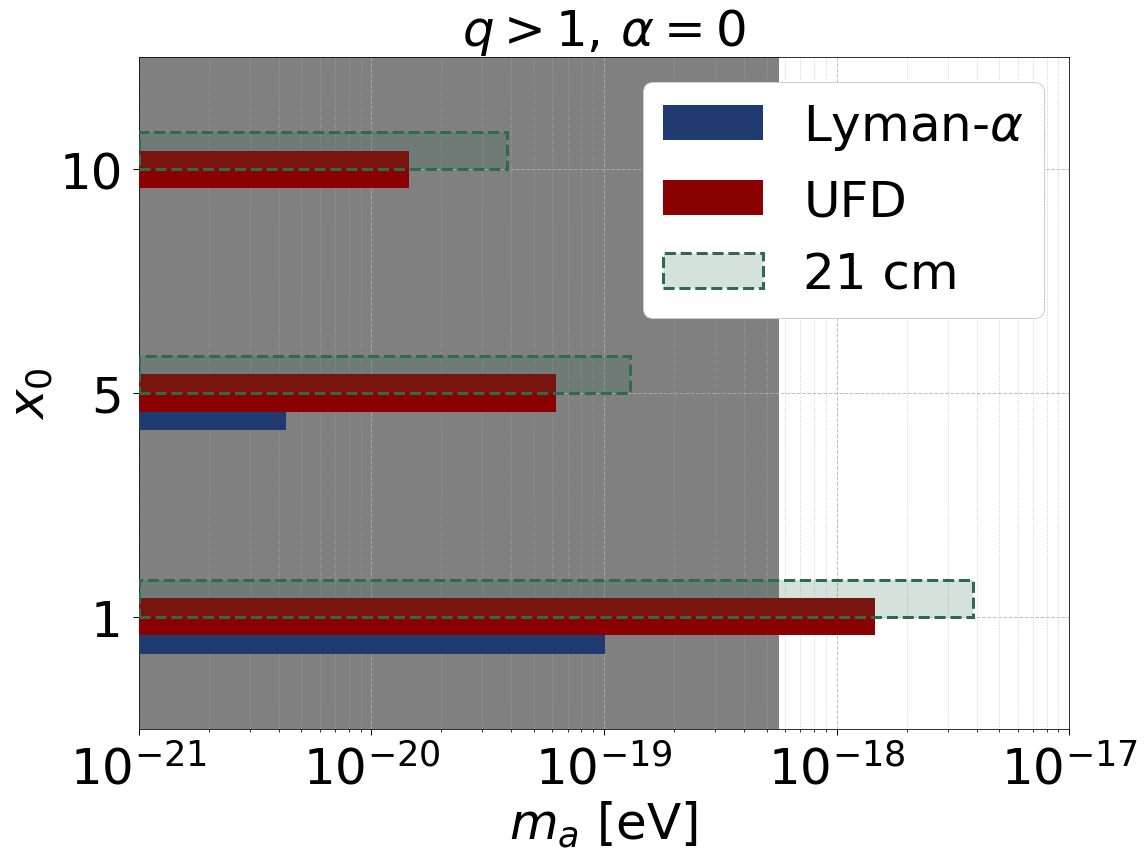} \quad
    \includegraphics[width=0.48\textwidth]{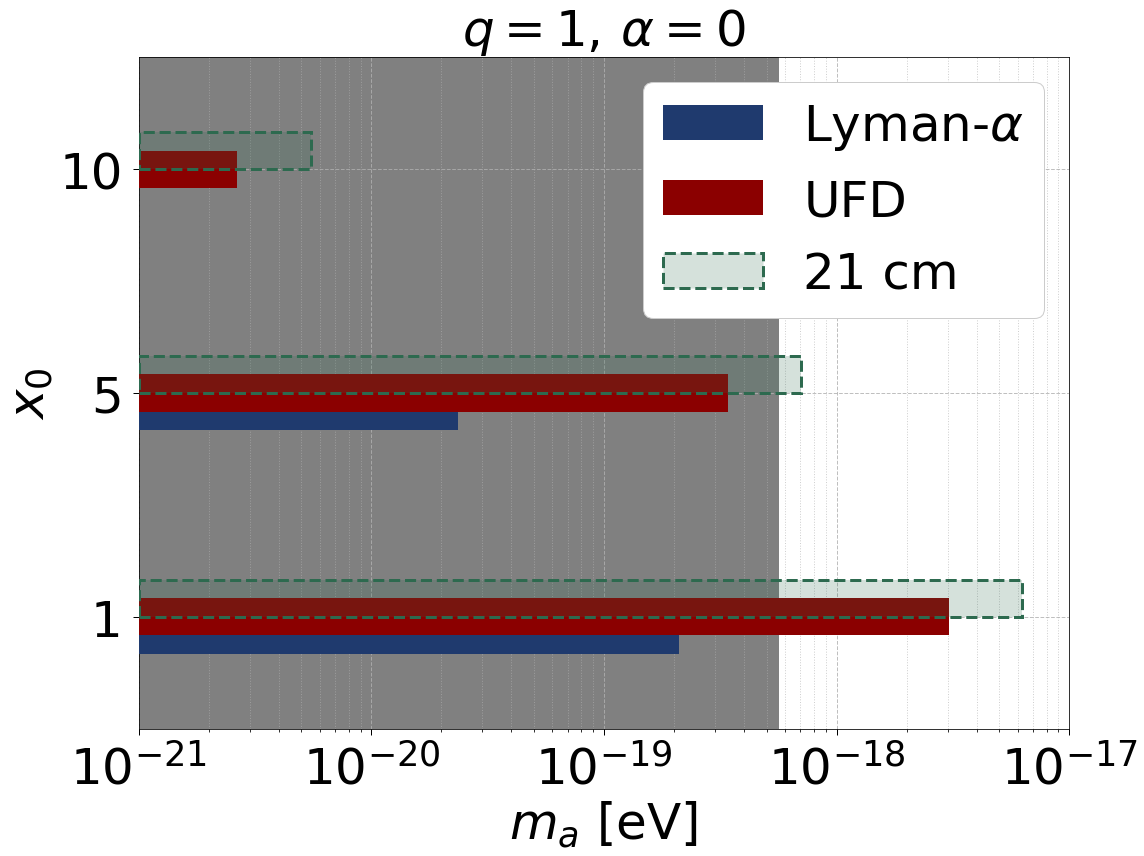} \\[3mm]
    \includegraphics[width=0.48\textwidth]
    {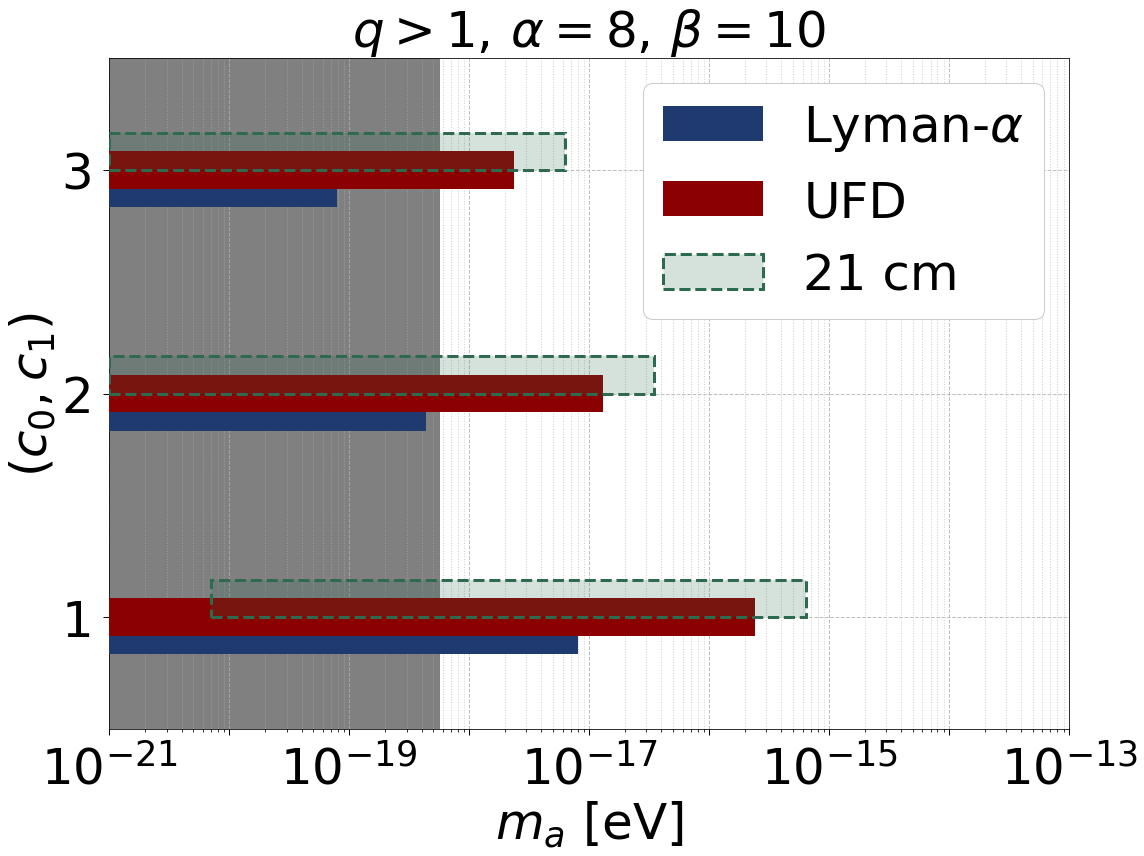}\quad
    \includegraphics[width=0.48\textwidth]{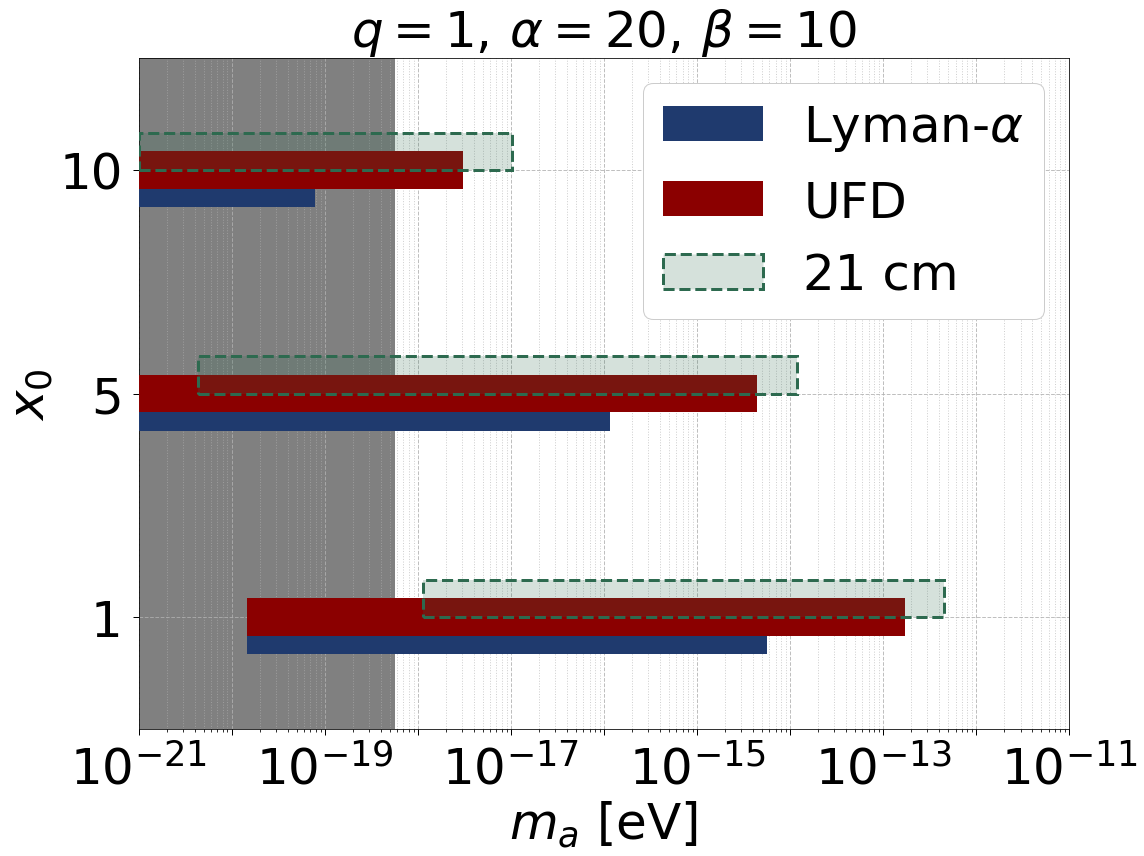}
    \caption{Bars show the range of $m_a$ excluded by the isocurvature constraints from Lyman-$\alpha$ (blue), UFD (red) and the sensitivity of 21~cm (light green) for different values of the IR cutoff parameter $x_0$ for $q>1$ and $\alpha=0$ (top-left), for $q=1$ and $\alpha=0$ (top-right) as well as $q=1$ and $\alpha = 20$ (bottom-right). The panel at the bottom-left shows the isocurvature constraints for different choices of the string density coefficients $(c_0, c_1)$. The cases are labeled as 1, 2, and 3 corresponding to (1, 0), (-0.81, 0.21) and (10, 1), respectively. The grey-shaded region indicates the lower bound on $m_a$ from free streaming, \cref{eq:ly-a_ma_bound_post} \cite{Liu:2024pjg}.}
    \label{fig:iso-bounds_ma}
\end{figure}

We summarize the resulting implications on the axion mass $m_a$ in \cref{fig:iso-bounds_ma}. The UFD limit from \cite{Graham:2024hah} improves substantially upon the Lyman-$\alpha$ constraint from \cite{Irsic2020}, typically tightening the lower limit on $m_a$ by over an order of magnitude. Further, the future 21~cm constraint from shock heating \cite{Irsic2020} will offer only a slight improvement in sensitivity compared to the UFD constraint. 

However, in general, isocurvature constraints give only a weak lower bound on $m_a$, typically less stringent than the constraint from free streaming. For $q>1,\,\alpha>0$ the main reason is the effect of the nonlinear transitions, which shift the peak of the spectrum to high momenta, c.f., \cref{fig:IRcutoff} (right), which suppresses the power spectrum. In these cases, the limits are almost independent of $\alpha,\,\beta$ and $x_0$. For $\alpha=0$, limits are weak due to the comparable large value of $k_{\rm min}$, as visible for the example in \cref{fig:IRcutoff} (left). For the cases $\alpha=0$ or $q=1$, the spectral peak is determined by the cutoff $k_{\rm min,*}$ and therefore limits are rather sensitive to the value of $x_0$. Physically, $1/k_{\rm min,*}$ corresponds to the largest axion wavelength the string network can radiate at $t_*$ and therefore $k_{\rm min,*} \sim H_*$. The prefactor relating $k_{\rm min,*}$ to the expansion rate is parametrized as $x_0\sqrt{\xi_*}$, where the factor $\sqrt{\xi_*}$ is motivated by the reduced inter-string distance for $\xi_* > 1$~\cite{Kim2024,Saikawa2024}. Numerically, we have $\sqrt{\xi_*} \approx 4.7$ in the low-$m_a$ region for our default assumption on the scaling behaviour. We consider $x_0 = 1,\,5,\,10$, covering roughly the range of $5 \lesssim x_0\sqrt{\xi_*} \lesssim 50$, motivated by simulations \cite{Kim2024,Gorghetto2021,Benabou2025,Saikawa2024}. 

The isocurvature constraint on $m_a$ is stronger than the one from free streaming for $q=1$ and $\alpha>0$ (see lower right panel of \cref{fig:iso-bounds_ma}). For $x_0=10$, we find $m_a > 3\times 10^{-18}$~eV from UFD. In this case, the bound is also strongly sensitive to the value of $x_0$ and can become very strong for $x_0<10$. The isocurvature constraint complements the free streaming bound and excludes the low mass region, such that the bound from SMBH superradiance of $m_a> 1.7\times 10^{-17}$ applies in this case, compare \cref{fig:strings_abundance_q1}.

We have also studied the dependence on the assumed scaling violation, as parametrised in \cref{eq:xi}. While the case $q>1,\alpha>0$ is insensitive to the choice of $x_0$, it does depend on the value of $\xi_*$, which affects the value of $k_{\rm peak}$ and therefore the isocurvature power $P_0$. We show this dependence in the bottom-left panel of \cref{fig:iso-bounds_ma}, where different choices of the parameters $c_0,c_1$ entering \cref{eq:xi} are adopted, where our default assumption corresponds to the case~2, and case~1 corresponds to the assumption of no scaling violations, i.e., $\xi=1$. In general, isocurvature constraints become stronger for smaller values of $\xi_*$. This also holds for $q>1,\alpha=0$ and $q=1$. However, while for $q>1,\alpha>0$ isocurvature constraints in the extreme case~1 provide substantially stronger lower bounds on $m_a$ than free streaming, in the other two scenarios the latter still dominates.

In summary, our updated isocurvature bounds on $m_a$ are significantly weaker than previous constraints \cite{Irsic2020,Feix2019,Feix2020}. This applies especially for temperature-dependent axion masses, but the bounds for $\alpha=0$ also become significantly weaker than previously thought, due to the cosmic string radiation production mechanism. The isocurvature bounds depend to some extent on the assumptions about string dynamics and are therefore subject to theoretical uncertainties.

%%%%%%%%%%%%%%%%%%%%%%%%%%%%%%%%%%%%%%%%%%%%%%%%%%%%%%%%%%%%%
\section{CMB tensor mode constraint for post-inflationary ALP dark matter}
\label{sec:tensor-modes}
%%%%%%%%%%%%%%%%%%%%%%%%%%%%%%%%%%%%%%%%%%%%%%%%%%%%%%%%%%%%%

For a given $f_a$, the post-inflation condition, $T_{\rm sym}(E_I,\epsilon) > f_a$ from \cref{eq:pre_post_condition}, can be interpreted as a lower bound on $E_I$. Since the dark matter production mechanism fixes $f_a$ for a given axion mass and ALP model parameters, this lower bound may be in tension with the upper bound on $E_I$ from CMB tensor modes, \cref{eq:EI_tensor_modes}. This provides an additional constraint on the viable parameter space, which depends on the reheating efficiency $\epsilon$. 

We denote the upper bound from tensor modes given in \cref{eq:EI_tensor_modes} by $E_I^{\rm max} = 1.4\times 10^{16}$~GeV. Using the definition of $T_{\rm sym}$ in \cref{eq:Tsym}, we obtain from the post-inflationary condition the upper bound
\begin{equation}\label{eq:fa_tensor-modes}
    f_a < \left\{
    \begin{array}{l@{\qquad}l}
      \sqrt{\frac{2}{3\pi}} (E_I^{\rm max})^2/M_{\rm pl}
      \approx
      7.5\times 10^{12}\,{\rm GeV} & \epsilon < \epsilon_{\rm crit}  \\[3mm]
      \epsilon E_I^{\rm max} \approx 1.4 \, \epsilon \times 10^{16}\,{\rm GeV}   & \epsilon > \epsilon_{\rm crit}
    \end{array}
    \right. \,,
\end{equation}
with
\begin{align}
  \epsilon_{\rm crit} \equiv 
  \sqrt{\frac{2}{3\pi}} \frac{E_I^{\rm max}}{M_{\rm pl}} \approx
   5.4\times 10^{-4} \,. 
\end{align}
For $\epsilon < \epsilon_{\rm crit}$, the bound is set by $T_{\rm GH}$ and is independent of $\epsilon$.

\begin{figure}[!t]
\centering
    \includegraphics[width=0.48\textwidth]{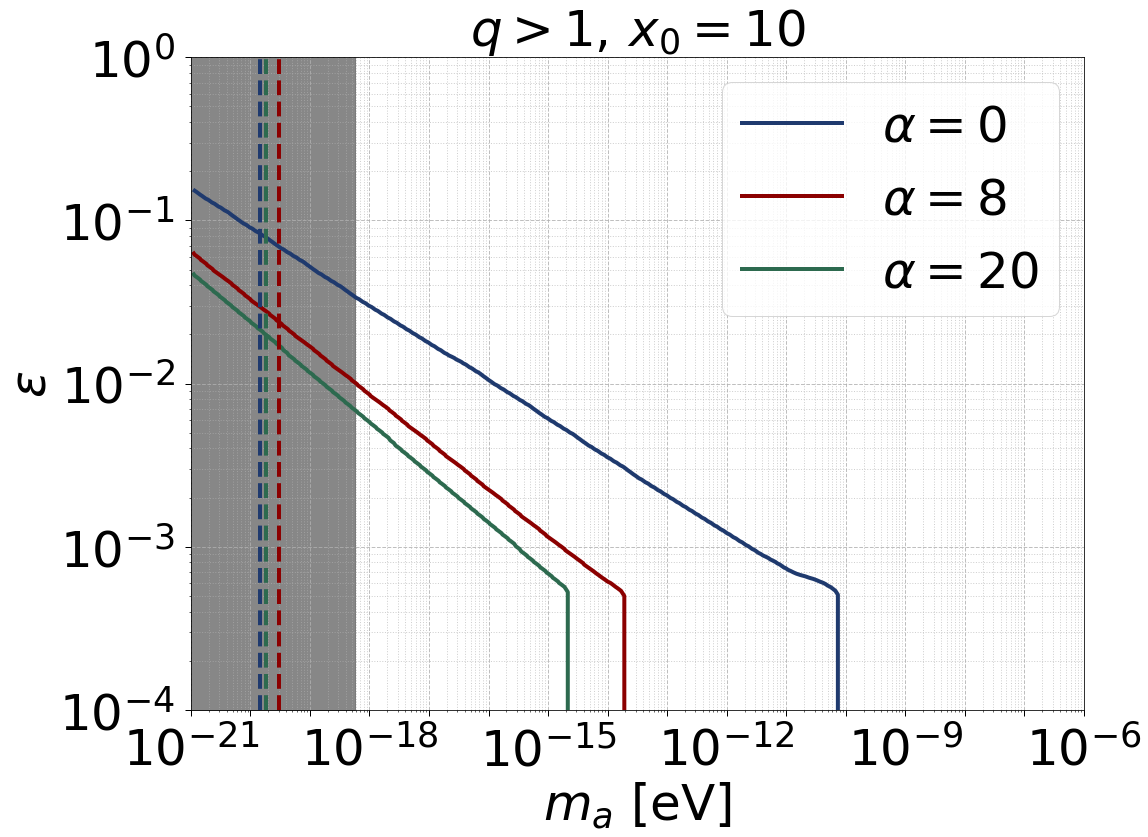}\quad
    \includegraphics[width=0.48\textwidth]{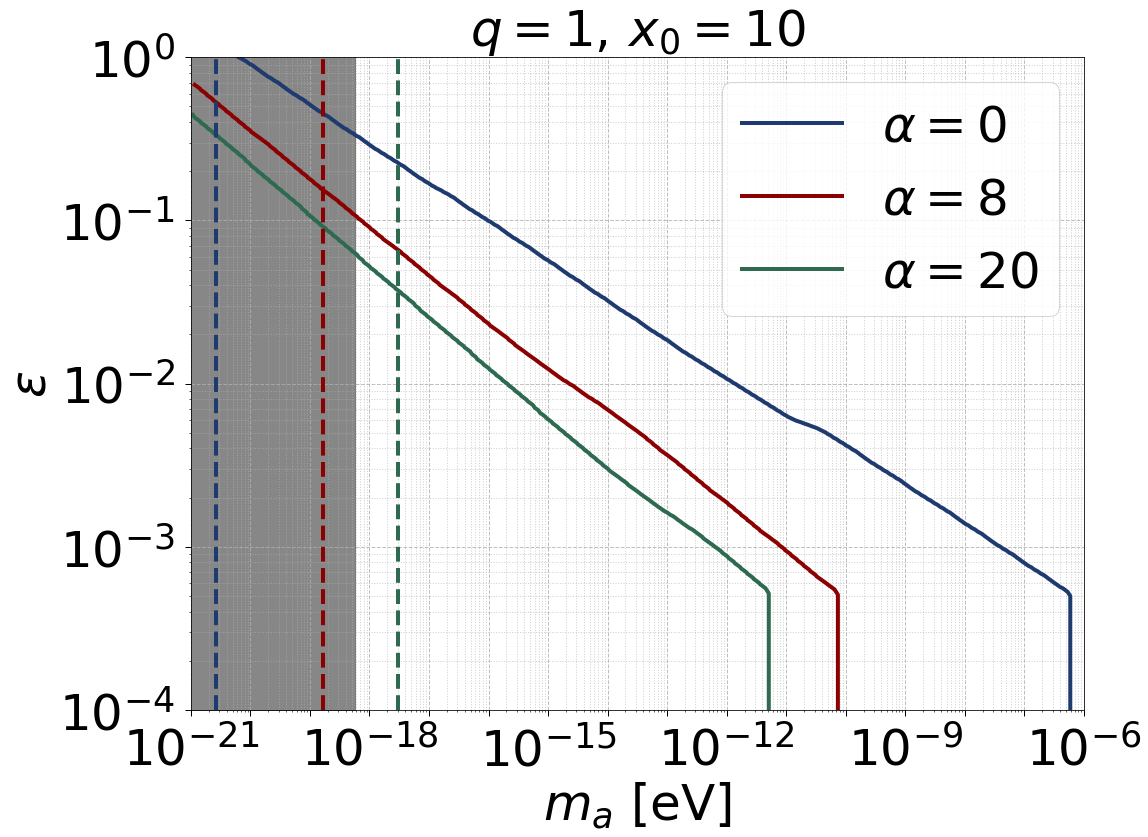} \\[3mm]
    \includegraphics[width=0.48\textwidth]{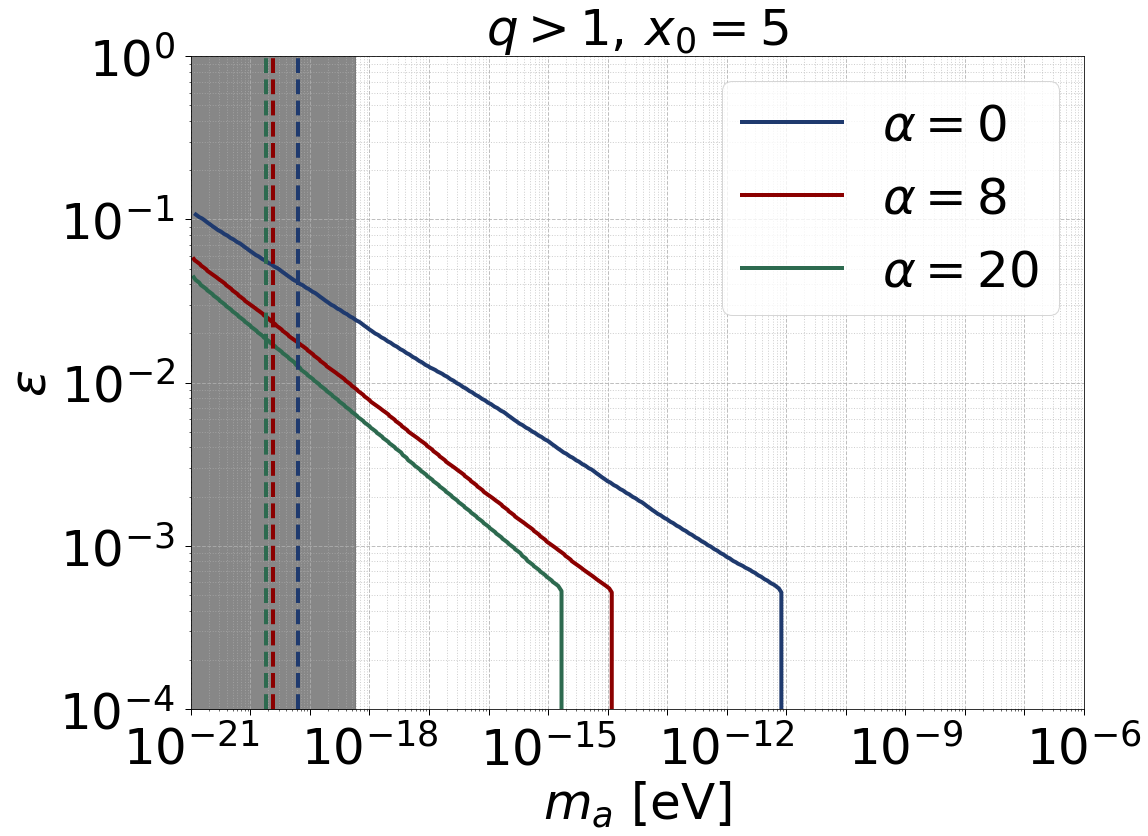}\quad
    \includegraphics[width=0.48\textwidth]{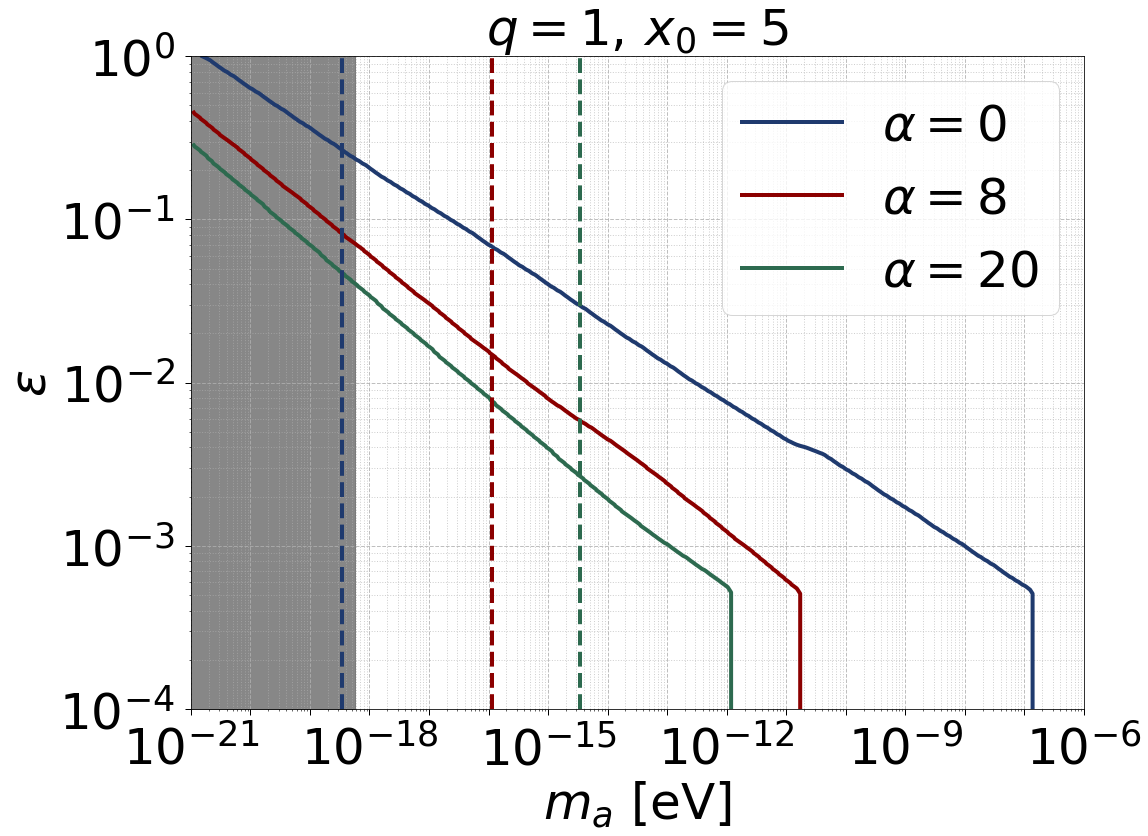}
    \caption{Solid lines show the lower bound on $m_a$ from CMB tensor modes as a function of the reheating efficiency for different choices of $\alpha$ and $\beta = 10$. Dashed vertical lines indicate the lower bound on $m_a$ from UFD isocurvature constraints from \cref{sec:iso-post}. Left (right) panels correspond to spectral index $q>1$ ($q=1$); for top (bottom) panels the momentum cutoff parameter $x_0$ is set to 10 (5). The grey-shaded region shows the bound from free streaming, \cref{eq:ly-a_ma_bound_post} \cite{Liu:2024pjg}. }
    \label{fig:tensor-modes}
\end{figure}

Now we can compare the upper bound from \cref{eq:fa_tensor-modes} with the values of $f_a$ required to obtain the full dark matter abundance. We illustrate this for a few $\epsilon$ values as horizontal lines in \cref{fig:strings_abundance_q3,fig:strings_abundance_q1}. The resulting lower bound on $m_a$ is shown as a function of the reheating efficiency $\epsilon$ in \cref{fig:tensor-modes} for a few model parameter choices. We observe rather strong bounds for low efficiencies and they become independent of $\epsilon$ for $\epsilon<\epsilon_{\rm crit}$. In turn, the constraints can be weakened if inflation models with large reheating efficiencies, $\epsilon \gtrsim 0.1$, are assumed. Then, the lower bound on $m_a$ from free streaming or isocurvature typically dominates, the latter shown as dashed vertical lines in \cref{fig:tensor-modes}.

Comparing the left and right panels in \cref{fig:tensor-modes}, we see that the bound is especially strong for $q=1$, which requires larger values of $f_a$ than $q>1$ for the same $m_a$. The same argument applies when comparing temperature-independent axion masses to cases with $\alpha > 0$. As the former requires larger $f_a$ than the latter, the tensor mode bound is stronger for $\alpha=0$. This nicely complements the isocurvature bounds, which show the opposite tendency and become stronger for large $\alpha$. For example, for $q=1$ or $\alpha = 0$ and $\epsilon\simeq 10^{-3}$, lower bounds can become as strong as $m_a \gtrsim 10^{-12}$~eV up to $10^{-7}$~eV, and therefore closing the gap between SMBH and stellar mass BH superradiance, or even extending beyond them.

In \cref{fig:tensor-modes} we fix $\beta = 10$. From 
\cref{fig:strings_abundance_q3,fig:strings_abundance_q1}, we see that changing $\beta$ from 10 to 0.1, may weaken the tensor mode bound on $m_a$ by up to one order of magnitude. We also note that tensor mode bounds are rather insensitive to the IR axion momentum cutoff parametrized by $x_0$.

%%%%%%%%%%%%%%%%%%%%%%%%%%%%%%%%%%%%%%%%%%%%%%%%%%%%%%%%%%%%%
\section{Pre-inflationary ALP dark matter}
\label{sec:pre}
%%%%%%%%%%%%%%%%%%%%%%%%%%%%%%%%%%%%%%%%%%%%%%%%%%%%%%%%%%%%%

We now turn to the pre-inflationary scenario and provide a brief review of axion dark matter production and isocurvature constraints. If the $U(1)$ symmetry breaks before the end of inflation, the axion acquires a single field value $\theta_0 f_a$ throughout the observable universe, with $\theta_0$ denoting the initial misalignment angle, which is random but spatially uniform. The axion behaves as cold dark matter once the Hubble rate becomes smaller than the mass and the field starts oscillating. The axion abundance is determined by the zero-momentum mode, that is, the homogeneous component of the field. As long as $\theta_0$ is not too close to the top of the potential around $\pi/2$, a rough estimate for the dark matter abundance is given by~\cite{Preskill:1982cy,Abbott:1982af,Dine:1982ah,Turner:1985si}:
\begin{equation}\label{eq:rho_pre}
    \rho(T_0) \approx m_a m_a(T_*) f_a^2 \theta_0^2 \left(\frac{a_*}{a_0}\right)^3 \,.
\end{equation}
For our numerical analysis, we compute the dark matter abundance by solving the equation of motion for the zero-momentum mode around the time $t_*$, accounting for the full cosine-shaped potential from \cref{eq:V}. We continue the numerical solution until (a) the field amplitude becomes small enough to justify the use of the quadratic approximation of the potential, and (b) the axion mass $m_a(T)$ becomes sufficiently larger than the Hubble rate $H(T)$, so that a WKB approximation is valid. 
We then match the numerical solution to the WKB ansatz, which we use until the axion mass reaches its zero-temperature value, allowing us to determine the final dark matter abundance today. Note that because of this approach, our results are insensitive to the precise definition of $T_*$ in \cref{eq:t_ast}, which serves only as a reference point for the numerical solution. In our work, we do not consider the fine-tuned case of $\theta_0 \approx \pi/2$ close to the top of the cosine potential, which would lead to additional phenomena (see Ref.~\cite{Arvanitaki:2019rax} for a discussion).

\begin{figure}[!t]
\centering
    \includegraphics[width=0.8\textwidth]{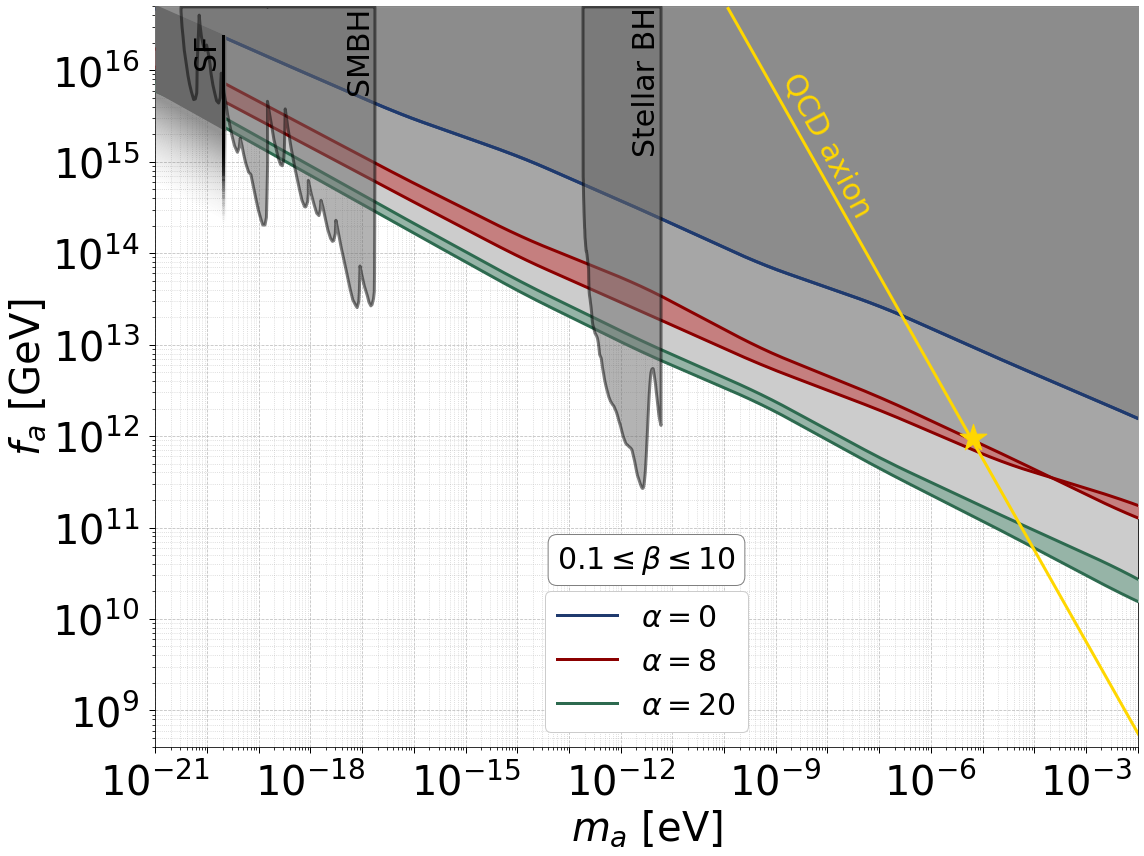} 
    \caption{Axion decay constant $f_a$ vs.\ axion mass $m_a$ in the pre-inflationary scenario, assuming that misalignment axions constitute all of dark matter. The width of the bands corresponds to changing $\beta$ from 0.1 to 10, with the lower (upper) boundary corresponding to 0.1 (10). The initial misalignment angle $\theta_0$ is fixed to 1. 
    Also shown are the lower bound on $m_a$ from structure formation (SF), \cref{eq:ly-a_ma_bound} \cite{Rogers:2020ltq}, regions excluded by BH superradiance \cite{Unal:2020jiy,Witte:2024drg}, and the QCD axion (yellow line) with the star marking where the QCD-like $m_a(T)$ gives the correct dark matter abundance.}
    \label{fig:pre_abundance}
\end{figure}

In \cref{fig:pre_abundance}, we show the value of $f_a$, which yields the correct dark matter abundance, as a function of $m_a$, for different values of $\alpha$ and $\beta$, with the latter having only a minor effect. Comparing these results with \cref{fig:strings_abundance_q1,fig:strings_abundance_q3}, we see that in the pre-inflationary scenario larger values of $f_a$ are required than for post-inflation. This follows from the enhancement factors for the string radiation abundance in \cref{eq:rho_mis_nonlin,eq:rho_mis_a0_q1}. For \cref{fig:pre_abundance}, we set the initial misalignment angle to $\theta_0=1$. Using \cref{eq:rho_pre}, we estimate the dependence on $\theta_0$ by applying \cref{eq:ma_T} for the temperature dependence of the axion mass and \cref{eq:t_ast} for the definition of $m_*$. Requiring constant $\rho(T_0)$ implies for fixed $m_a$
\begin{equation}
    \label{eq:scaling_theta0}
    f_a  \theta_0^{\frac{4\alpha + 16}{3\alpha+16}} \approx const.\,, 
\end{equation}
where we have neglected the temperature dependence of the relativistic degrees of freedom. For $\alpha=0$, the product $f_a\theta_0$ remains constant to very good accuracy. For $\alpha>0$, we have checked that \cref{eq:scaling_theta0} agrees with the numerical result within 10\%. Hence, larger values of $f_a$ than shown in \cref{fig:pre_abundance} are required for $\theta_0 < 1$.

We see also from \cref{fig:pre_abundance} that black hole superradiance severely affects the parameter space in the pre-inflationary case. In addition to the stellar-mass BH exclusion region around $m_a\sim 10^{-12}$~eV, supermassive BHs provide a lower bound of $m_a > 1.7\times 10^{-17}$~eV, which holds for $\alpha \lesssim 8$. For $\alpha \approx 20$, two tiny allowed regions remain between $10^{-19}$ and $10^{-18}$~eV. These bounds will be even tighter for $\theta_0 < 1$.

\bigskip

If the $U(1)$ symmetry breaks before the end of inflation, the axion already exists during inflation and, like any light scalar present at that time, acquires quantum fluctuations. These lead to isocurvature perturbations with the amplitude set by the Gibbons-Hawking temperature from \cref{eq:T_GH}. For further discussion, see, e.g., 
\cite{Lyth:1991ub,Beltran:2006sq,Hertzberg:2008wr,Arvanitaki:2019rax} and \cref{app:iso_pre}. Their power spectrum is given by
\begin{equation}\label{eq:Delta_pre}
    \Delta^2_a(k) = \frac{H_I^2}{\pi^2 \theta_0^2 f_a^2} \,.
\end{equation}
In contrast to the post-inflationary case discussed above, the isocurvature spectrum in the pre-inflationary scenario is approximately scale-invariant, similar to the adiabatic fluctuations. Planck CMB data constrain such a scale-free isocurvature power spectrum with  $\beta_{\rm iso} = f_{\rm iso}^2/(1+f_{\rm iso}^2) < 0.038$ at 95\%~CL~\cite{Planck:2018jri}. This corresponds to
\begin{equation}
        f_{\rm iso} = \frac{H_I}{\pi \theta_0 f_a \sqrt{A_s}} 
    < 0.2 \,.
    \label{eq:fiso_pre_planck}
\end{equation}

\begin{figure}[!t]
\centering
    \includegraphics[width=0.6\textwidth]{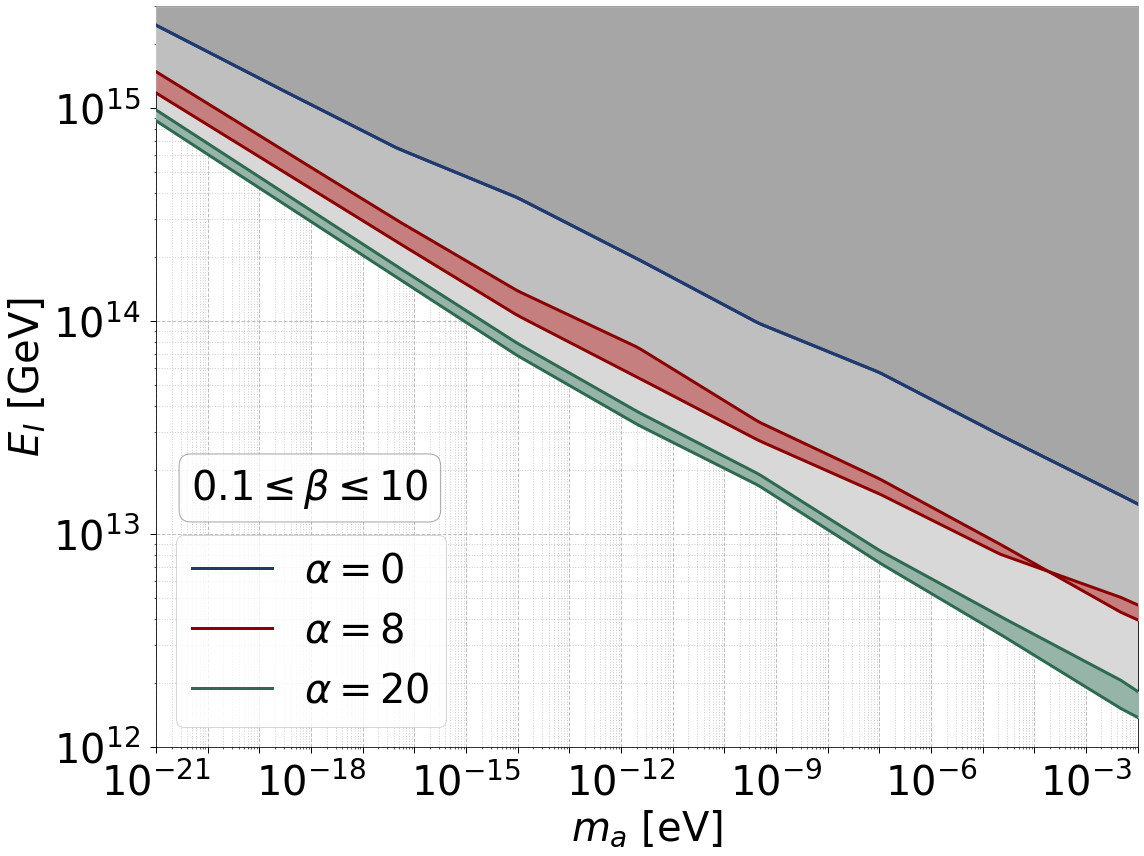}
    \caption{Upper bound on the inflationary energy scale $E_I$ in the pre-inflationary case
    %, see \cref{eq:fiso_pre_planck} \cite{Planck:2018jri}, 
    vs. axion mass $m_a$. Shaded regions above the curves overproduce isocurvature fluctuations. The width of the bands corresponds to changing $\beta$, with the lower (upper) boundary corresponding to 0.1 (10). The initial misalignment angle $\theta_0$ is fixed to 1 and $f_a$ is determined by requiring full axion dark matter abundance. }
    \label{fig:E_I}
\end{figure}

Using the relation \cref{eq:HI_EI} between the Hubble rate during inflation $H_I$, and the energy density during inflation, \cref{eq:fiso_pre_planck} leads to an upper bound on the energy scale of inflation $E_I$: 
\begin{equation}\label{eq:bound_EI}
    E_I < 3.5\times 10^{14}\,{\rm GeV} \sqrt{\frac{\theta_0 f_a}{10^{15}\, {\rm GeV}}} \,.
\end{equation}
This bound is shown in \cref{fig:E_I} as a function of $m_a$ for different values of $\alpha$ and $\beta$. Note that for a temperature-independent axion mass, this upper bound is independent of $\theta_0$, as it depends on the product $\theta_0f_a$ fixed by the dark matter abundance. For $\alpha>0$, the bounds show a weak dependence on $\theta_0$ and can easily be rescaled by using \cref{eq:scaling_theta0,eq:bound_EI}.\footnote{For completeness, we mention that the authors of Ref.~\cite{Caputo:2023ikd} derived a lower bound on $E_I$ for pre-inflation ALP dark matter based on isocurvature fluctuations, $E_I \gtrsim 140 \,{\rm TeV}\,\sqrt{m_a/{\rm eV}}$. However, this applies to a very different region of parameter space than considered here.}

Comparing \cref{fig:E_I} with the bound from CMB tensor modes in \cref{eq:EI_tensor_modes}, we see that the upper bound from isocurvature perturbations in the pre-inflationary axion dark matter case is stronger. A possible observation of tensor modes in future CMB observations \cite{Abazajian2022} would then essentially exclude pre-inflationary ALP dark matter from the misalignment mechanism.\footnote{This argument can be avoided if the quartic coupling $\lambda$ is very small, and the complex scalar field is coupled to the Ricci scalar via a term $\xi R |\phi|^2$. This can lead to an effective $f_a$ during inflation, larger than the one after inflation, which would suppress the isocurvature perturbations generated during inflation and allow for observable tensor modes \cite{Linde:1991km,Folkerts:2013tua,Fairbairn:2014zta,Graham:2025iwx}.
Here we assume generic couplings $\lambda \sim 1$ (as we also do for the post-inflationary case), such that the value of $f_a$ is the same during and after inflation.}

%%%%%%%%%%%%%%%%%%%%%%%%%%%%%%%%%%%%%%%%%%%%%%%%%%%%%%%%%%%%%
\section{Post- versus pre-inflation and the energy scale of inflation}
\label{sec:post-vs-pre}
%%%%%%%%%%%%%%%%%%%%%%%%%%%%%%%%%%%%%%%%%%%%%%%%%%%%%%%%%%%%%

\begin{figure}[!t]
\centering
    \includegraphics[width=0.48\textwidth]{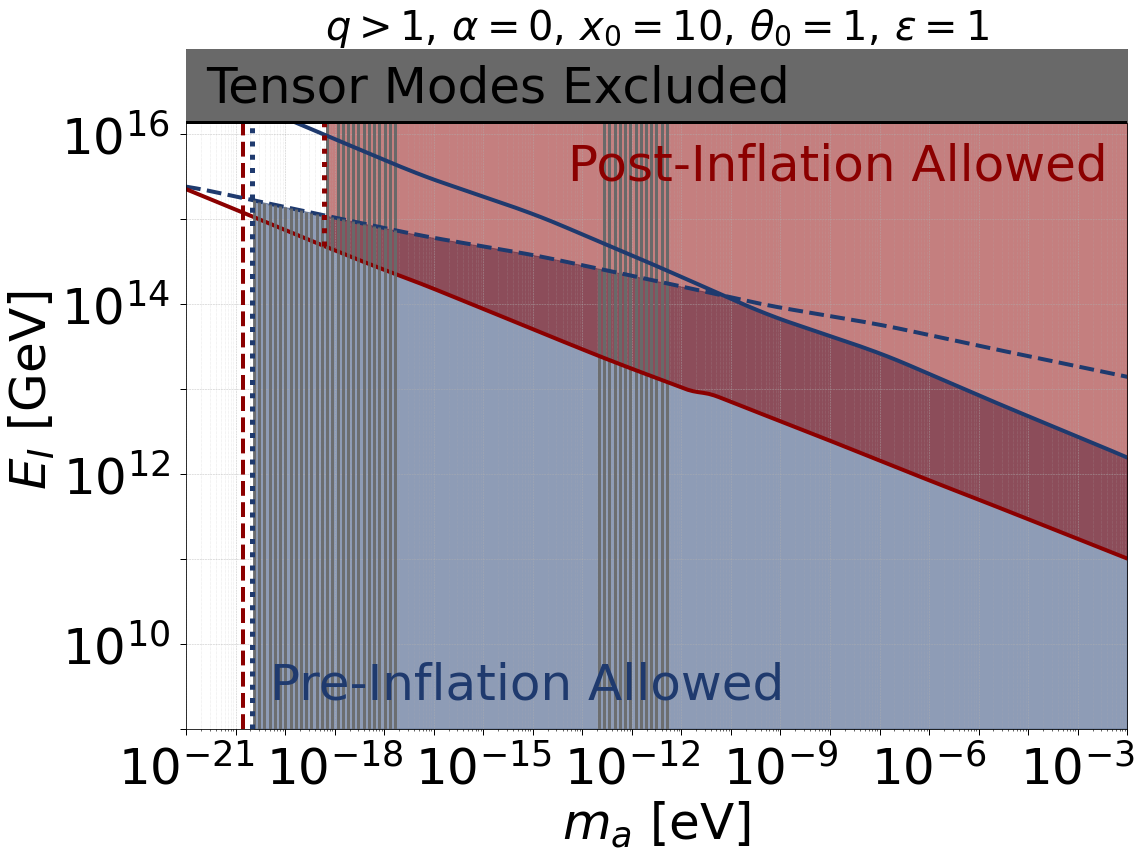}\quad
    \includegraphics[width=0.48\textwidth]{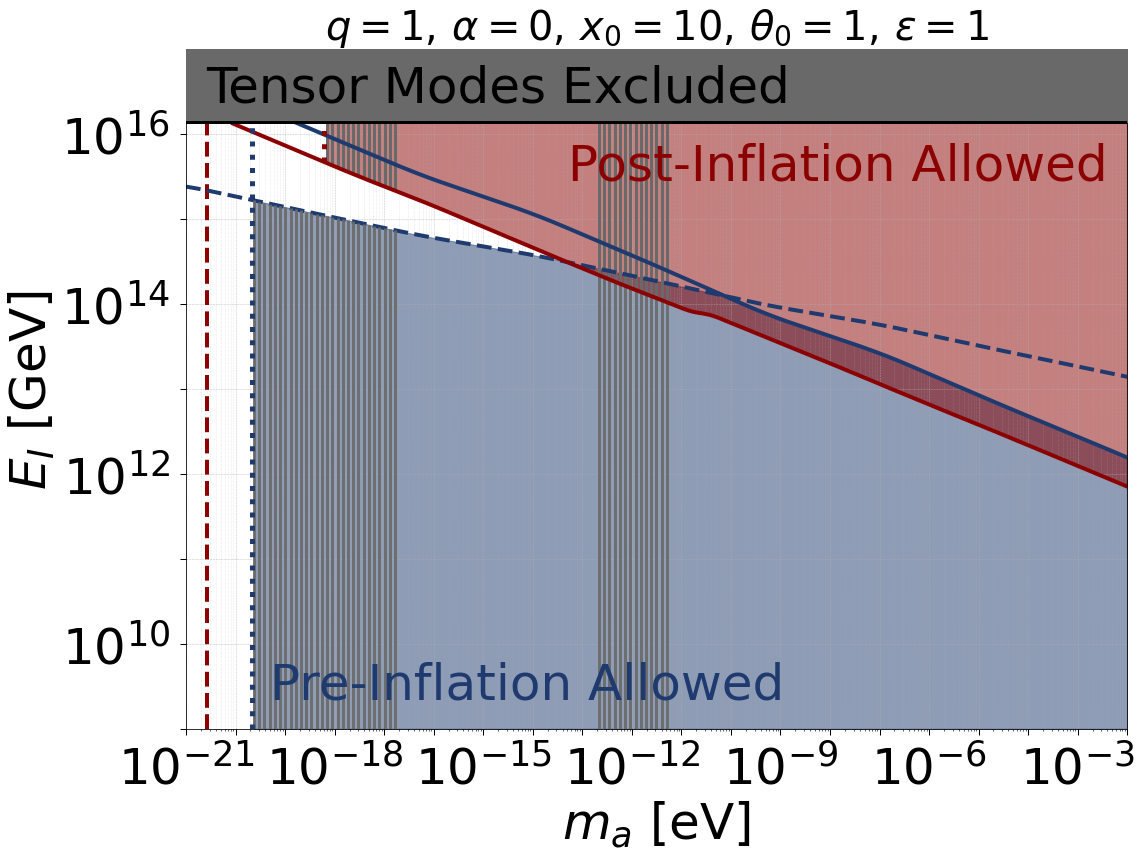} \\[3mm]
    \includegraphics[width=0.48\textwidth]{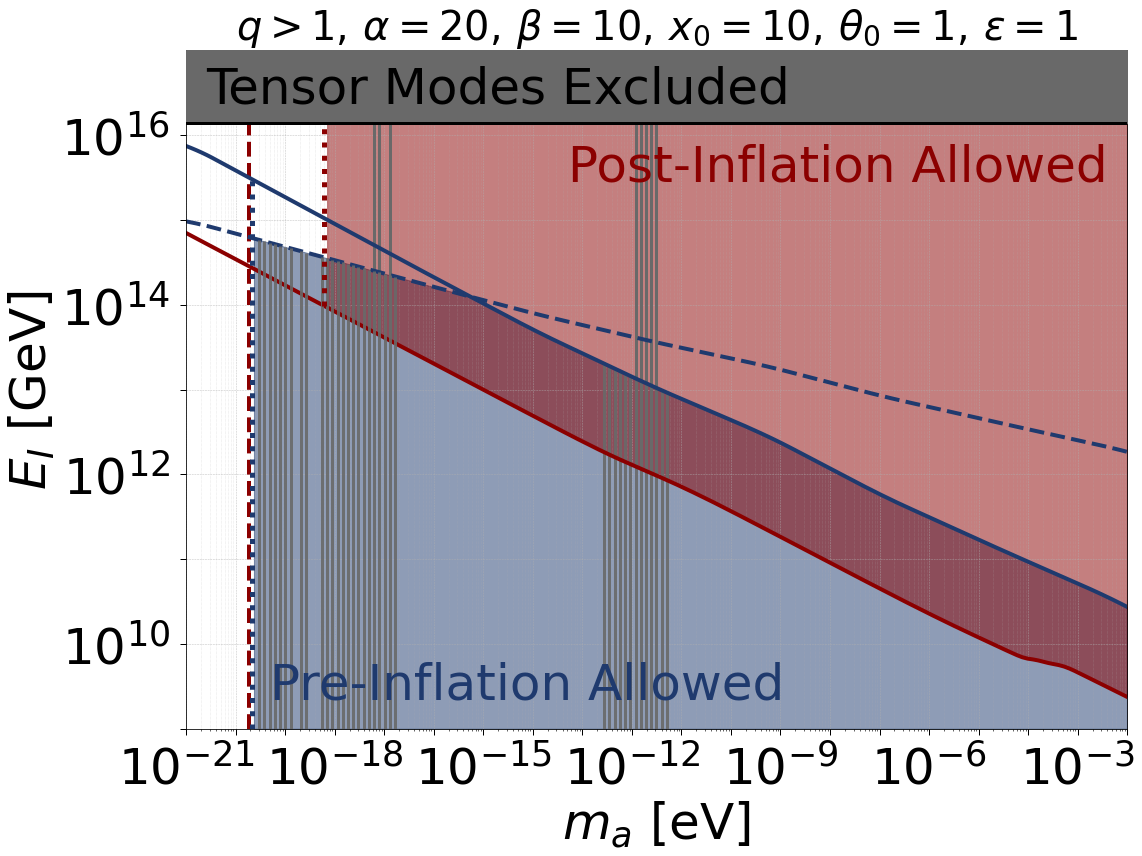}\quad
    \includegraphics[width=0.48\textwidth]{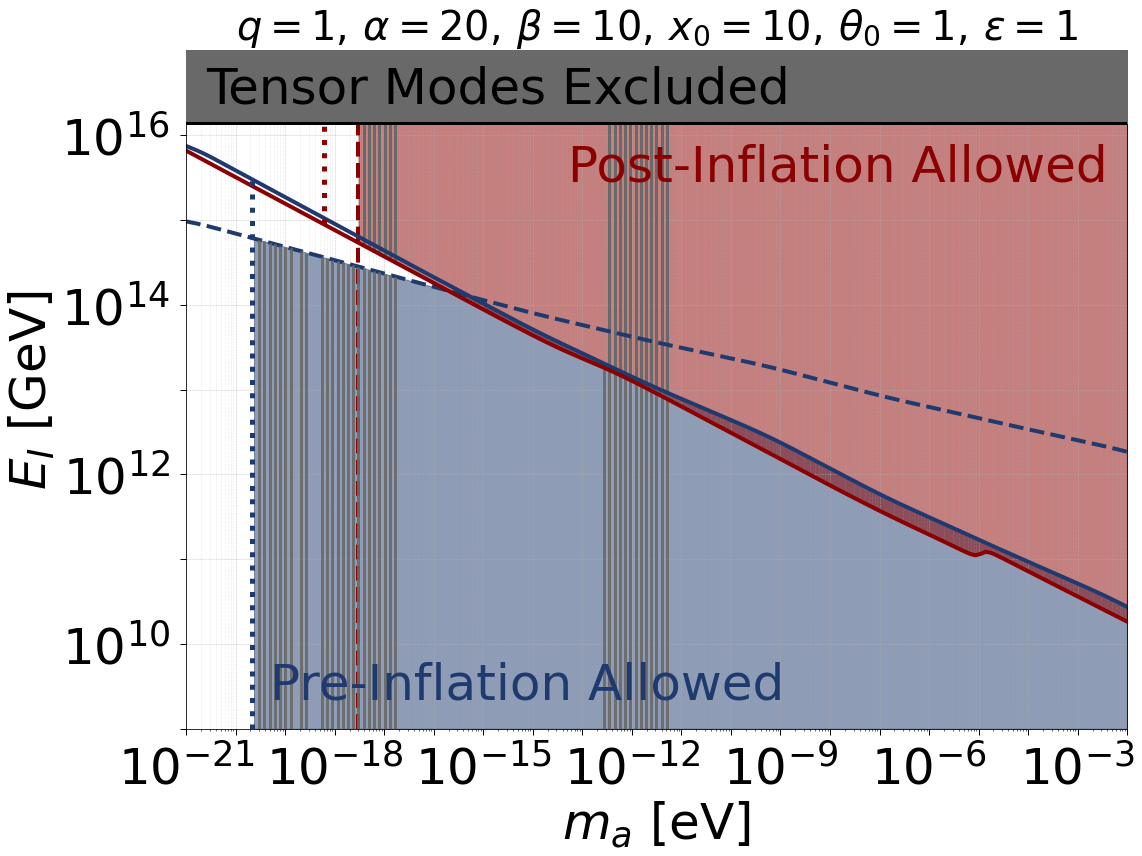}
    \caption{Allowed regions in the plane of axion mass and energy scale of inflation for post-inflation (red) and pre-inflation (blue) for the parameter choices indicated in the title of each panel. The region above the red-solid (below the blue-solid) curve corresponds to the condition $f_a^{\rm post} < T_{\rm sym}(E_I,\,\epsilon)$ ($f_a^{\rm pre} > T_{\rm sym}(E_I,\,\epsilon))$. The region to the left of the red-dashed line (above the blue-dashed line) is excluded by isocurvature constraints in the post-\mbox{(pre-)inflationary} case. The region to the left of the red-(blue-)dotted line is excluded by structure formation in the post-\mbox{(pre-)}inflationary case. Excluded regions from CMB tensor modes are marked with grey shading. Grey vertically-hatched regions are excluded by black hole superradiance. Upper (lower) panels correspond to temperature-independent(-dependent) axion masses. Left (right) panels assume a spectral index of $q>1$ ($q=1$) for the axion emission spectrum in post-inflation.}
    \label{fig:EI-ma_1}
\end{figure}

\begin{figure}[!t]
\centering
    \includegraphics[width=0.48\textwidth]{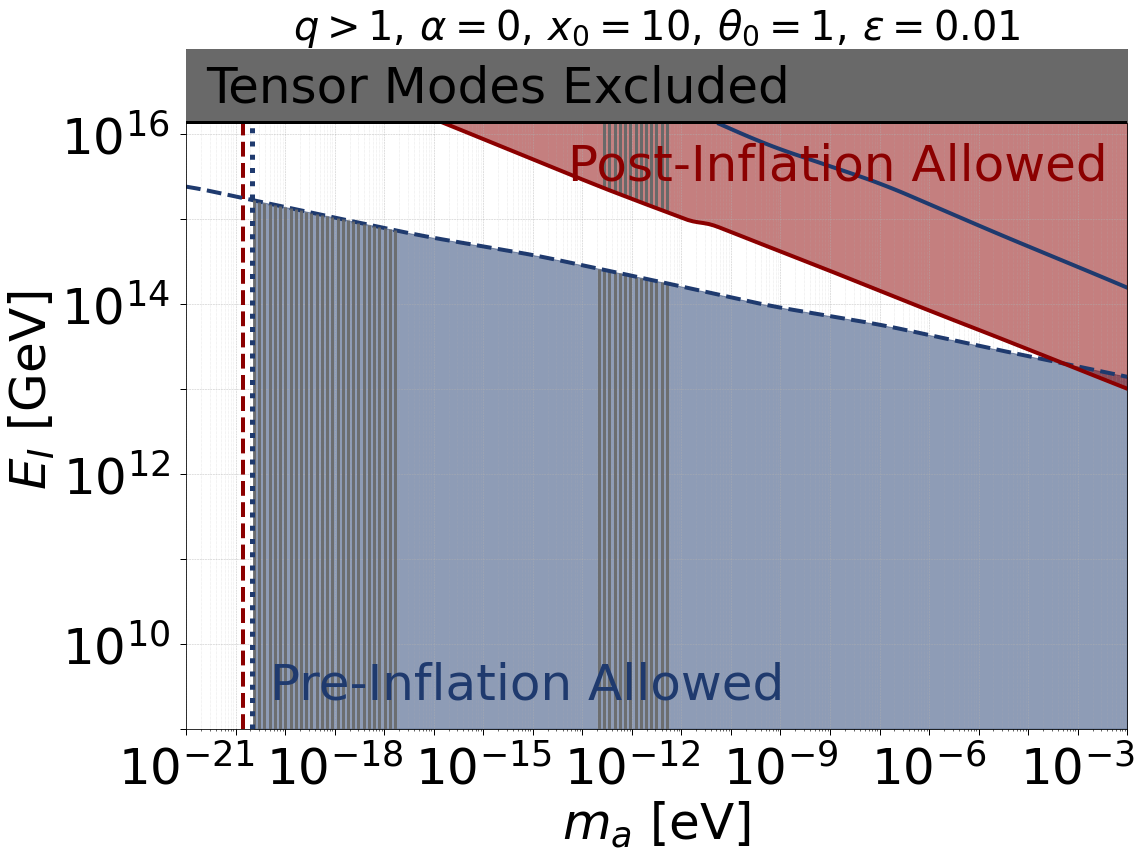}\quad
    \includegraphics[width=0.48\textwidth]{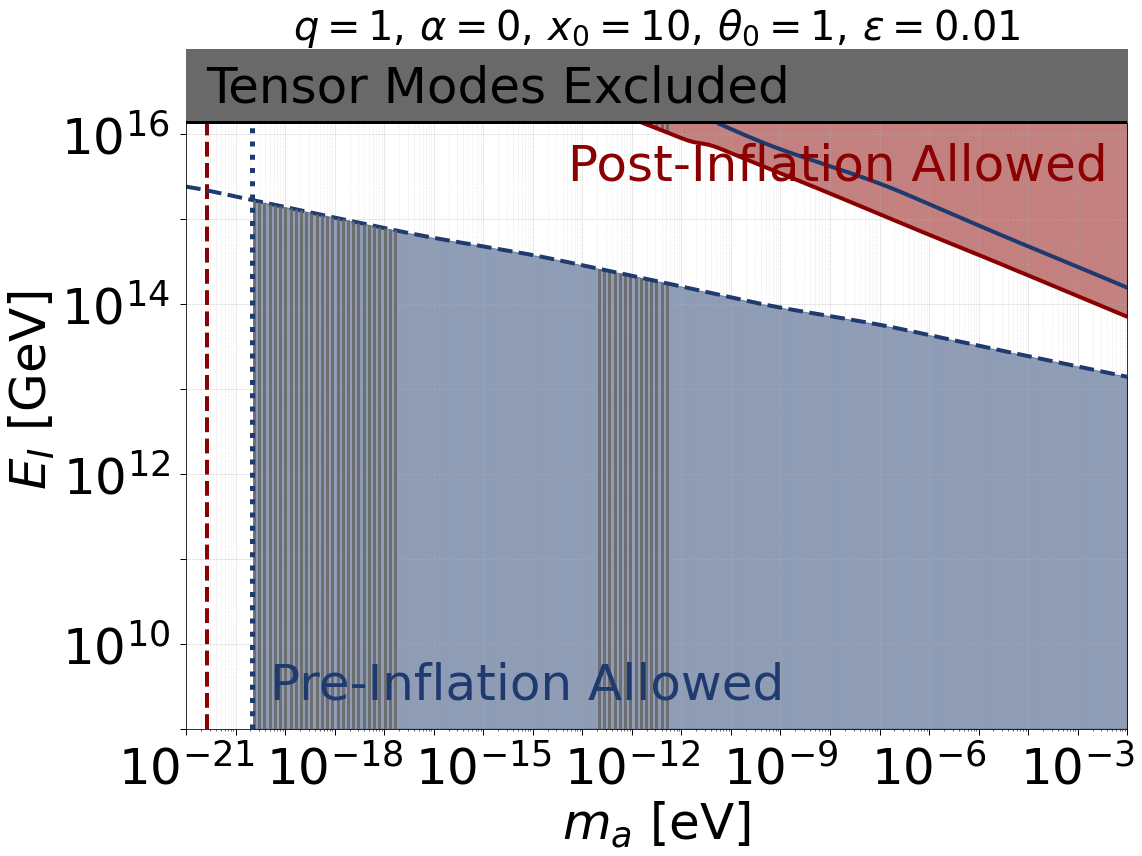} \\[3mm]
    \includegraphics[width=0.48\textwidth]{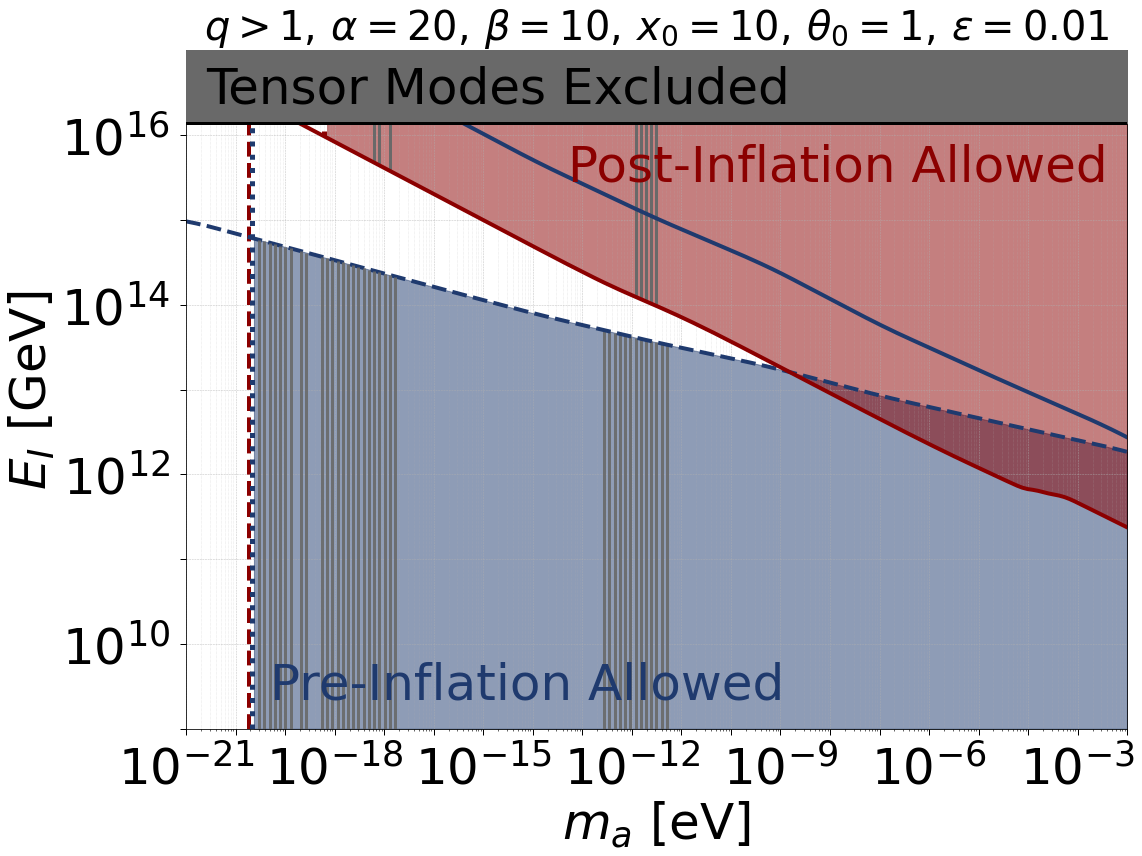}\quad
    \includegraphics[width=0.48\textwidth]{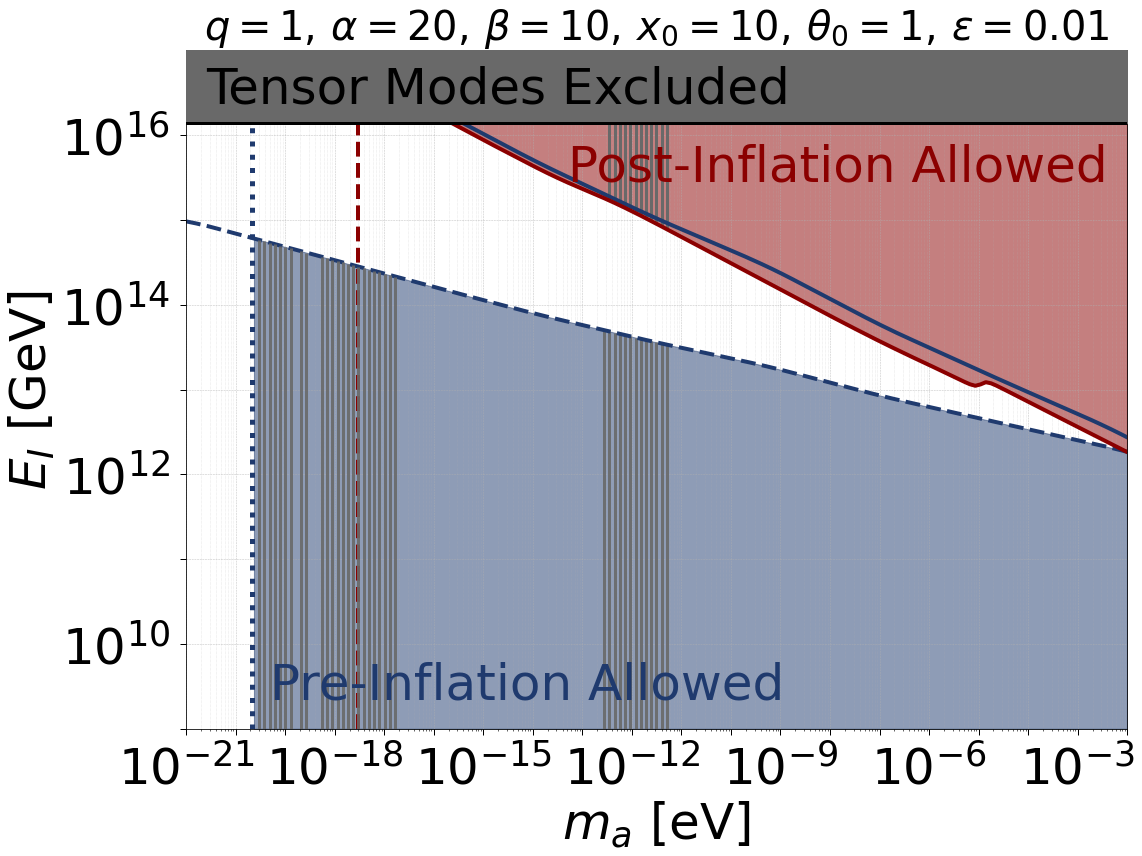}
    \caption{Same as \cref{fig:EI-ma_1}, but assuming a reheating efficiency of $\epsilon = 0.01$.}
    \label{fig:EI-ma_2}
\end{figure}

\begin{figure}[!t]
\centering

   \includegraphics[width=0.48\textwidth]{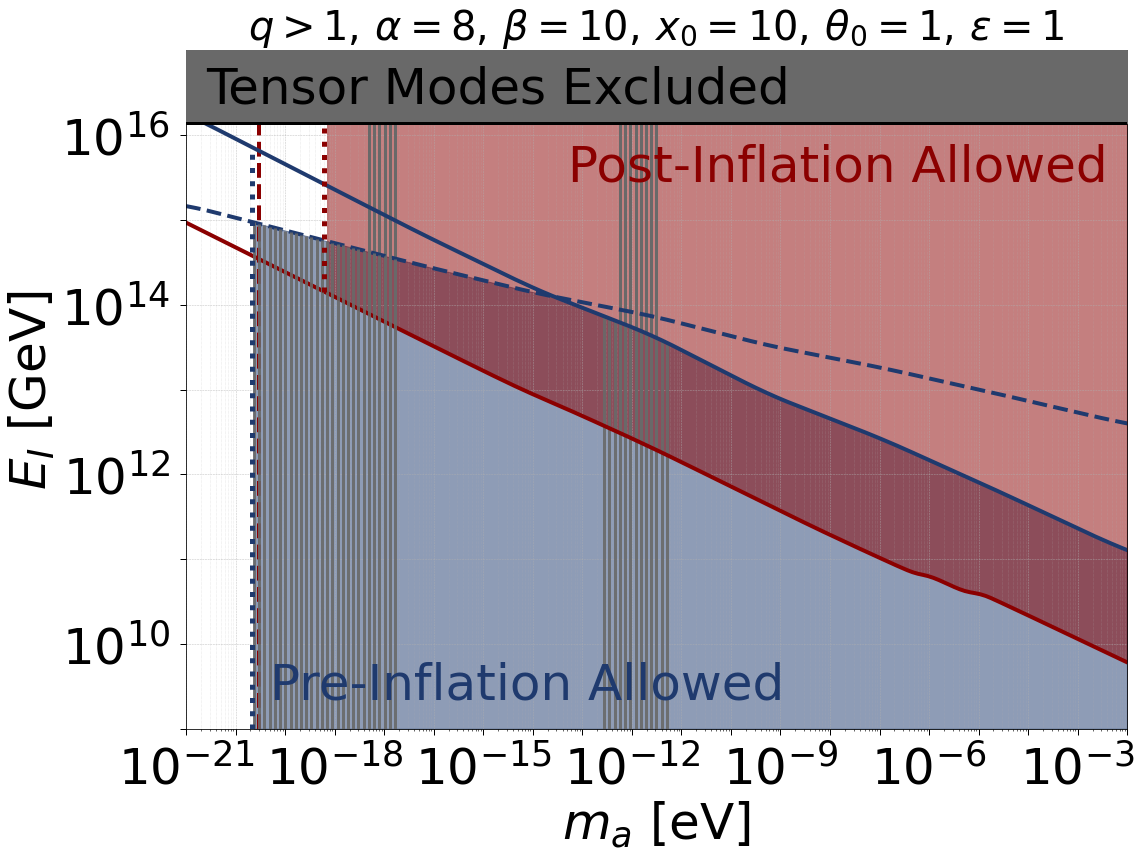}\quad
    \includegraphics[width=0.48\textwidth]{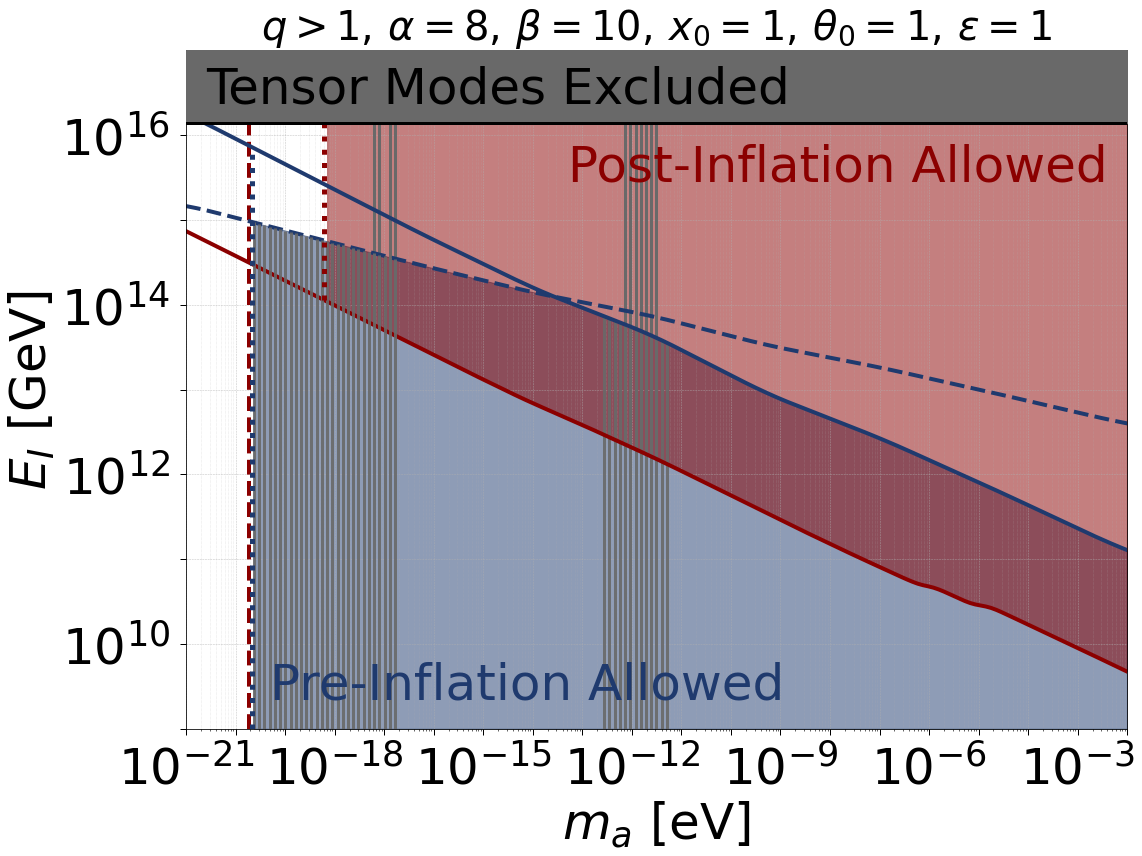} \\[3mm]
    \includegraphics[width=0.48\textwidth]{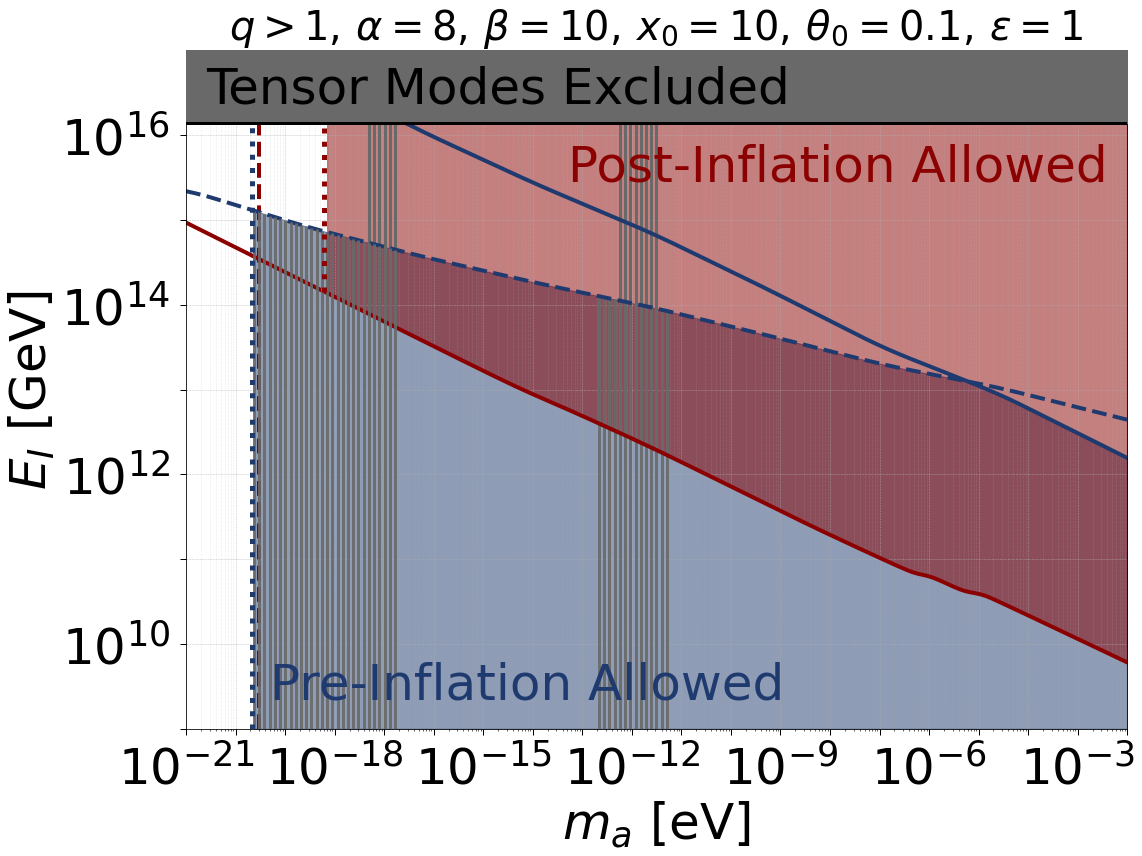}\quad
    \includegraphics[width=0.48\textwidth]{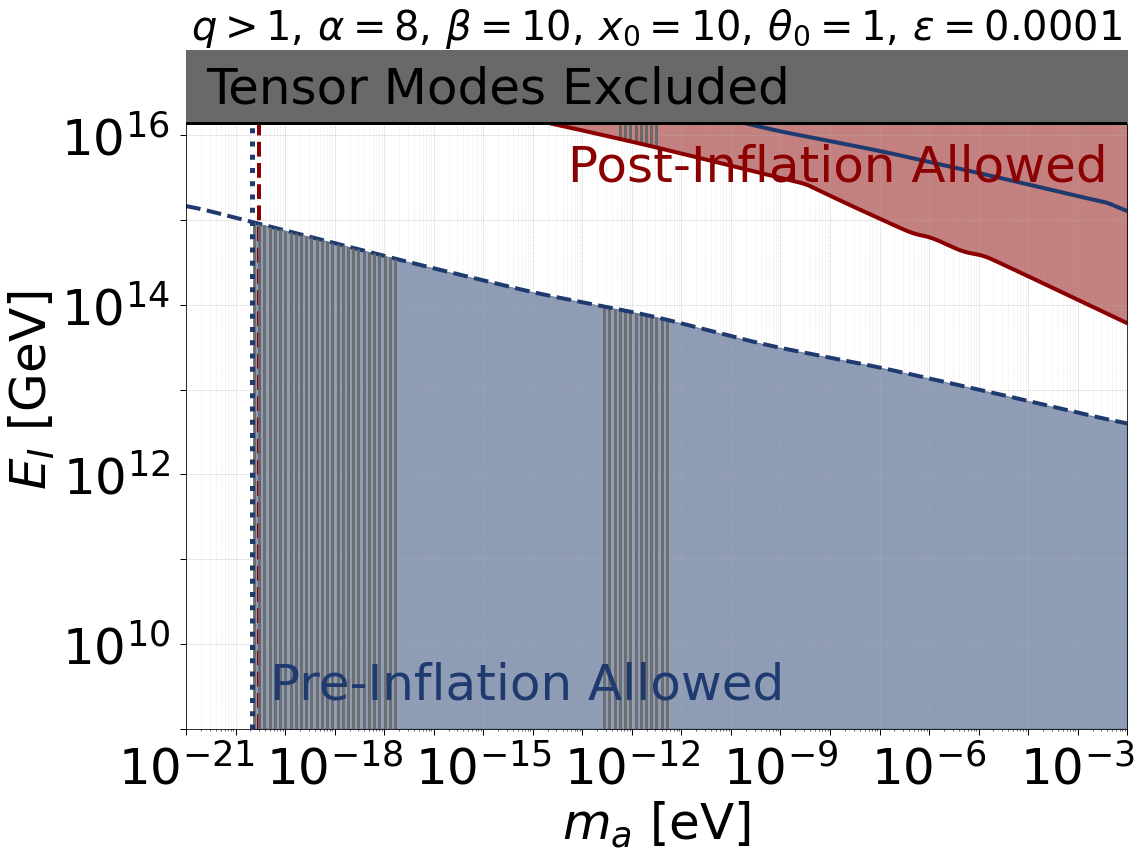}
    \caption{Same as \cref{fig:EI-ma_1} for $q>1,\, \alpha=8 \text{ and }\beta=10$ (all panels). We take the upper-left panel as reference and change the IR cutoff parameter for the post-inflationary axion spectrum $x_0 = 10 \to 1$ (upper-right), the initial misalignment angle for pre-inflation $\theta_0=1\to 0.1$ (lower-left), and the reheating efficiency $\epsilon=1\to 10^{-4}$ (lower-right).}
    \label{fig:EI-ma_3}
\end{figure}

Let us now combine the above parameter space constraints for post- and pre-inflationary ALP dark matter and derive allowed regions for the energy scale of inflation $E_I$ as a function of the axion mass. Some examples are shown in \cref{fig:EI-ma_1,fig:EI-ma_2,fig:EI-ma_3}, where the various panels illustrate the dependence of the regions on the model parameters and the uncertainties on the axion emission spectrum, which follow from the results discussed in the previous sections. The regions shaded in red (blue) in these figures show the allowed parameter space for post-inflation (pre-inflation). Dark red indicates overlapping regions where both scenarios can provide the full dark matter abundance, while white corresponds to regions of the parameter space where neither can.

As discussed above, the $f_a$ required for the full dark matter abundance in the pre-inflationary case ($f_a^{\rm pre}$) is in most cases larger than the one for post-inflation ($f_a^{\rm post}$). Hence, taking into account the post-/pre-inflation condition in \cref{eq:pre_post_condition}, for a fixed axion mass and given $T_{\rm sym}(E_I,\,\epsilon)$, we expect the following cases: 
\begin{equation}\label{eq:pre-post-cases}
    \begin{split}
    &T_{\rm sym} < f_a^{\rm post} < f_a^{\rm pre}: \quad\text{only pre-inflation,}\\ 
    &f_a^{\rm post} < T_{\rm sym} < f_a^{\rm pre}: \quad\text{pre- and post-inflation possible,}\\
    &f_a^{\rm post} < f_a^{\rm pre} < T_{\rm sym}: \quad\text{only post-inflation.}  
    \end{split}
\end{equation}
For the $E_I$ values below the solid blue curves in \cref{fig:EI-ma_1,fig:EI-ma_2,fig:EI-ma_3}, the pre-inflation condition $T_{\rm sym} < f_a^{\rm pre}$ is satisfied, whereas the regions above the red curves fulfil the post-inflation condition $T_{\rm sym} > f_a^{\rm post}$. The region between the red and blue curves corresponds to the intermediate situation from \cref{eq:pre-post-cases}, where both scenarios can provide the full dark matter abundance for these values of $E_I$. 

The dashed lines in \cref{fig:EI-ma_1,fig:EI-ma_2,fig:EI-ma_3} indicate the isocurvature constraints, providing a lower bound on $m_a$ for post-inflation (we show the constraint from UFD~\cite{Graham:2024hah}) and an upper bound on $E_I$ from CMB in pre-inflation~\cite{Planck:2018jri}, see~\cref{eq:bound_EI}. The latter can, depending on parameters, supersede the upper bound from the pre-inflation condition $T_{\rm sym} < f_a^{\rm pre}$. The union of these two bounds determines the upper bound on $E_I$ in pre-inflation, which is typically significantly stronger than the CMB tensor mode bound in \cref{eq:EI_tensor_modes}, see the discussion in \cref{sec:pre}. The dotted blue (red) lines show the lower limit on the axion mass $m_a$ from structure formation (free streaming) in pre-inflation (post-inflation), see \cref{eq:ly-a_ma_bound} (\cref{eq:ly-a_ma_bound_post}). The CMB tensor mode bound in the post-inflationary scenario discussed in \cref{sec:tensor-modes} is visible in the panels, where the red-solid curve crosses the CMB bound at higher $m_a$ values than the lower limits from isocurvature and free streaming. The vertical grey-hatched regions in the figures are excluded by BH superradiance from \cite{Unal:2020jiy,Witte:2024drg}. These are independent of $E_I$, but can differ for post- versus pre-inflation, as the required values of $f_a^{\rm post}$ and $f_a^{\rm pre}$ are different for fixed $m_a$. 

Upper panels of \cref{fig:EI-ma_1} correspond to a temperature-independent axion mass, whereas lower panels show a strong temperature dependence with $\alpha = 20$.
Typically, the superradiance bounds from SMBHs provide the lower bound $m_a > 1.7\times 10^{-17}$~eV in the pre-inflationary case. However, for models with strong temperature dependence of the axion mass, islands of allowed regions below this bound appear. For post-inflation, a combination of BHSR, free streaming and isocurvature constraints can also provide a robust lower bound $m_a > 1.7\times 10^{-17}$~eV in most cases, except for $q>1$ and large $\alpha$. Whether isocurvature or free streaming bounds are stronger, depends sensitively on the model parameters. The former dominates for $q=1$ and large $\alpha$, as seen in the lower-right panel. In most cases, stellar mass BHs exclude the region around $m_a\sim 10^{-12}$~eV. 

\Cref{fig:EI-ma_2} corresponds to the same assumptions as \cref{fig:EI-ma_1}, except that the reheating efficiency is chosen as $\epsilon=0.01$ instead of 1. We see that the post-inflationary regions shift to larger axion masses, leading to the strongest lower bound on the mass coming from CMB tensor modes in the upper and lower-right panels. Similarly, the solid blue curve delimiting the pre-inflation condition is also shifted towards larger $m_a$ and $E_I$, such that the isocurvature constraint dominates the upper bound on $E_I$ over the whole considered range for $m_a$. Furthermore, for a broad range of $m_a$ values, a region in $E_I$ appears, where neither post- nor pre-inflation can provide the full dark matter abundance. 

In \cref{fig:EI-ma_3}, we illustrate the effect of changing other parameters, where the upper-left panel serves as our benchmark configuration. The upper-right panel shows the impact of changing the lower IR cutoff for the axion emission spectrum from $x_0 = 10$ to $1$. While the isocurvature bound in post-inflation remains unchanged due to the nonlinear transitions for $q>1,\,\alpha>0$, it has an effect on the required $f_a$ and therefore changes the BHSR constraints. The lower-left panel shows the impact of reducing the initial misalignment angle in pre-inflation from $\theta_0 = 1$ to $0.1$, which basically shifts the solid blue curve to larger $m_a$ and $E_I$, whereas the isocurvature constraint is only weakly affected, making the latter the dominating upper bound on $E_I$. Finally, the lower-right panel shows a very small value for the reheating efficiency of $\epsilon=10^{-4}$. This shifts both the red and blue solid lines to larger $m_a$ and $E_I$. For pre-inflation, this has a similar effect as decreasing $\theta_0$, whereas for post-inflation the allowed region is pushed to high values of $E_I$, close to the CMB tensor mode bound, leading to a rather strong lower bound on $m_a$ and predicting high values for the energy scale of inflation. We note the kink in the red solid curve visible in the lower-right panel, which marks the transition  of $T_{\rm sym}$ from $T_{\rm max}$ to $T_{\rm GH}$, see \cref{eq:Tsym}.

%%%%%%%%%%%%%%%%%%%%%%%%%%%%%%%%%%%%%%%%%%%%%%
\section{Summary and discussion}
\label{sec:conclusion}
%%%%%%%%%%%%%%%%%%%%%%%%%%%%%%%%%%%%%%%%%%%%%%

In this work, we assume that axions, pseudo-Goldstone bosons from a global $U(1)$ symmetry spontaneously broken at a high scale $f_a$, provide the full dark matter abundance in the universe. We remain agnostic about the mechanism of explicit symmetry breaking and parametrize the axion mass as a temperature-dependent power-law, $m_a(T) \propto T^{-\alpha/2}$ with $\alpha \ge 0$, cut off at the zero-temperature value $m_a$. The case $\alpha = 0$ corresponds to a temperature-independent axion mass. We focus exclusively on gravitational phenomena and do not consider any direct coupling to the Standard Model (except for the temperature-dependence of the mass, which may emerge from interactions with the Standard Model plasma). Our main guiding principle is the dark matter abundance, allowing us to determine $f_a$ for a given zero-temperature axion mass $m_a$. We study both the post- and pre-inflationary symmetry breaking scenarios and derive lower bounds on the axion mass using various observables. 

In \cref{sec:ALP_DM_post,sec:iso-post,sec:tensor-modes}, we focus on the post-inflationary case, with the main results summarized in \cref{fig:strings_abundance_q3,fig:strings_abundance_q1}. We review the dominating axion production mechanism from the cosmic string network during the scaling regime. We update previous isocurvature constraints \cite{Feix2019,Feix2020,Irsic2020} by considering this production mechanism, with limits from Lyman-$\alpha$ observations \cite{Irsic2020} and ultra-faint dwarfs (UFDs) \cite{Graham:2024hah}, as well as forecasts for the sensitivity of future 21~cm observations from the shock heating of baryonic gas due to dark matter halos \cite{Irsic2020}. Typically, we find much weaker constraints than obtained previously \cite{Irsic2020,Feix2019,Feix2020}, because of a suppressed isocurvature amplitude due to the cosmic string production mechanism and nonlinear transitions shifting the spectral peak to higher axion momenta. Even the very strong isocurvature bounds from UFDs \cite{Graham:2024hah} lead in most cases to weaker bounds on $m_a$ than free streaming effects in structure formation.

In \cref{sec:tensor-modes}, we show that the upper bound on the energy scale of inflation $E_I$ from CMB tensor bounds yields a complementary lower bound on the axion mass
by requiring that the post-inflationary condition between $f_a$ and  $E_I$ is fulfilled and that the full dark matter abundance is reproduced. This bound depends sensitively on the temperature dependence of the axion mass, on the axion emission spectral index $q$, as well as on the reheating efficiency. It strengthens with decreasing $\alpha$ and is most stringent for a temperature-independent mass, while being stronger for $q=1$ than for $q>1$. The tensor mode bound becomes particularly strong for inefficient reheating, $\epsilon \lesssim 0.01$, and saturates for $\epsilon < 5.4\times 10^{-4}$ at values ranging from $m_a > 2\times 10^{-15}$~eV  for $\alpha=20$ and $q>1$ to $m_a > 5\times 10^{-7}$~eV for $\alpha=0$ and $q=1$. 

After briefly reviewing the pre-inflationary scenario in \cref{sec:pre}, we compare both cases in \cref{sec:post-vs-pre}, highlighting allowed regions in the $m_a-E_I$ plane. Combining the bounds discussed above with black hole superradiance constraints and structure formation limits from Lyman-$\alpha$ power spectrum observations \cite{Rogers:2020ltq} (including the effect of free streaming in the post-inflationary case \cite{Amin:2022nlh,Liu:2024pjg}), we can summarize our findings regarding the question posed in the title of this paper as follows:
\begin{itemize}
\item 
For the pre-inflationary scenario, supermassive black hole superradiance (SMBH SR) \cite{Unal:2020jiy} leads to the lower bound $m_a> 1.7\times 10^{-17}$~eV. It holds for initial misalignment angles $\theta_0 < 1$ and irrespective of the axion mass temperature dependence, except for $\alpha \gtrsim 20$ small allowed islands appear 
for $10^{-19}\,{\rm eV} \lesssim m_a \lesssim 10^{-18}\,{\rm eV} $.
\item In the post-inflationary scenario, the SMBH SR bound 
$m_a> 1.7\times 10^{-17}$~eV applies in the case of a temperature-independent axion mass.
\item For temperature dependent axion masses in the post-inflationary case, a key systematic uncertainty is the spectral index of the axion emission spectrum $q$. For $q=1$, the combination of free streaming,  isocurvature, and SMBH SR leads again to the bound 
$m_a> 1.7\times 10^{-17}$~eV. 
\item This bound can be avoided in post-inflation if $q>1$ and $\alpha \gtrsim 8$. In this case, isocurvature bounds are strongly suppressed due to nonlinear transitions and allowed regions appear below the SMBH SR bound, although this statement depends to some extent on assumptions about the string dynamics. The constraint is then set by the free streaming effect on the matter power spectrum: $m_a > 5.6\times 10^{-19}$~eV \cite{Rogers:2020ltq,Liu:2024pjg}. 
\item 
Significantly stronger bounds on $m_a$ are obtained from CMB tensor modes in the post-inflationary scenario if the efficiency of reheating is not too large, i.e., $\epsilon<0.01$, which is the case in generic models of reheating, e.g., \cite{Mukaida:2015ria}, see \cref{fig:tensor-modes}. 
\end{itemize}
The limits derived here complement other potential constraints, such as those from stellar heating of ultra-faint dwarf galaxies, which may also apply in the $10^{-19} -10^{-17}$~eV range~\cite{Marsh:2018zyw,Dalal:2022rmp,Teodori:2025rul,May:2025ppj,Chiang:2021uvt,Yang:2025bae}, or from recurrent axinovae \cite{Fox:2023xgx}.

Finally, we note some caveats and limitations of our analysis. Our results rely on the assumed dark matter production mechanism, axion radiation from strings in post-inflation and standard misalignment in pre-inflation, as well as their theoretical and systematic uncertainties. In particular, uncertainties of the axion emission spectrum in the post-inflationary scenario, remain a big challenge. Due to the huge dynamical range relevant for the string network evolution, results of numerical simulations need to be extrapolated by many orders of magnitude. We recap below the relevant parameters and their impact on our results:
An important assumption is the logarithmic scaling violation of the string network, parametrized by the parameters $c_0,c_1$ in \cref{eq:xi}. We have studied a wide range for their values including also the case of no scaling violations. These parameters modify the amplitude of the power spectrum and can have an impact on the isocurvature constraints. We note however, that the lower bounds on $m_a$ summarized above are conservative; depending on the choices of $c_0$ and $c_1$, isocurvature bounds may supersede the SMBH SR bound in some cases. A similar comment applies to the parameter $x_0$ entering the low-$k$ cut-off, \cref{eq:kmin}: small values can lead to strong isocurvature constraints for $q=1$, but our benchmark choice $x_0 = 10$ gives conservative mass bounds. Finally, the coefficients $c_v, c_m, c_n$ introduced in \cref{sec:ALP_DM_post} encode details of the axion energy spectrum and number density, including transient effects and higher-mode contributions. While they can modify the overall axion abundance, their precise values do not impact the resulting mass limits quoted above.
One assumption we make is that the spectral peak during the second nonlinear transition follows the virialized momentum $k_{\mathrm{vir}}$, see \cref{eq:k_peak}, as found in simulations of the QCD axion in ref.~\cite{Gorghetto:2024vnp}. 
The spectral index $r$ determines the shape of the emitted axion spectrum after the second nonlinear transition, see \cref{eq:rho_DM_2nd-NL}. Variations of its value leave the inferred mass constraints unchanged. Importantly, the second nonlinear transition only modifies the isocurvature constraints for temperature-dependent axion masses with $q>1$, where it does not dominate the resulting mass bounds.

Finally, we mention that if additional axion production mechanisms (for example from the decay of the string-domain wall network) or even other dark matter particles contribute significantly to the total dark matter abundance, the phenomenology changes: smaller $f_a$ are required, relaxing the bounds discussed here. Also, for alternative axion production mechanisms, see e.g.,~\cite{Co:2019jts,Chang:2019tvx, Eroncel:2022vjg,Arvanitaki:2009fg,DiLuzio:2021gos,Bodas:2025eca,Harigaya:2022pjd,Eroncel:2025qlk}, bounds may change and additional phenomena may become relevant.

\appendix
%%%%%%%%%%%%%%%%%%%%%%%%%%%%%%%%%%%%%%%%%%%%%%%%%%%%%%%%%%%%%%%%
\section{Exact scaling of string network} \label{sec:constant-string-density}
%%%%%%%%%%%%%%%%%%%%%%%%%%%%%%%%%%%%%%%%%%%%%%%%%%%%%%%%%%%%%%%

\begin{figure}[htbp]
    \includegraphics[width=0.48\textwidth]{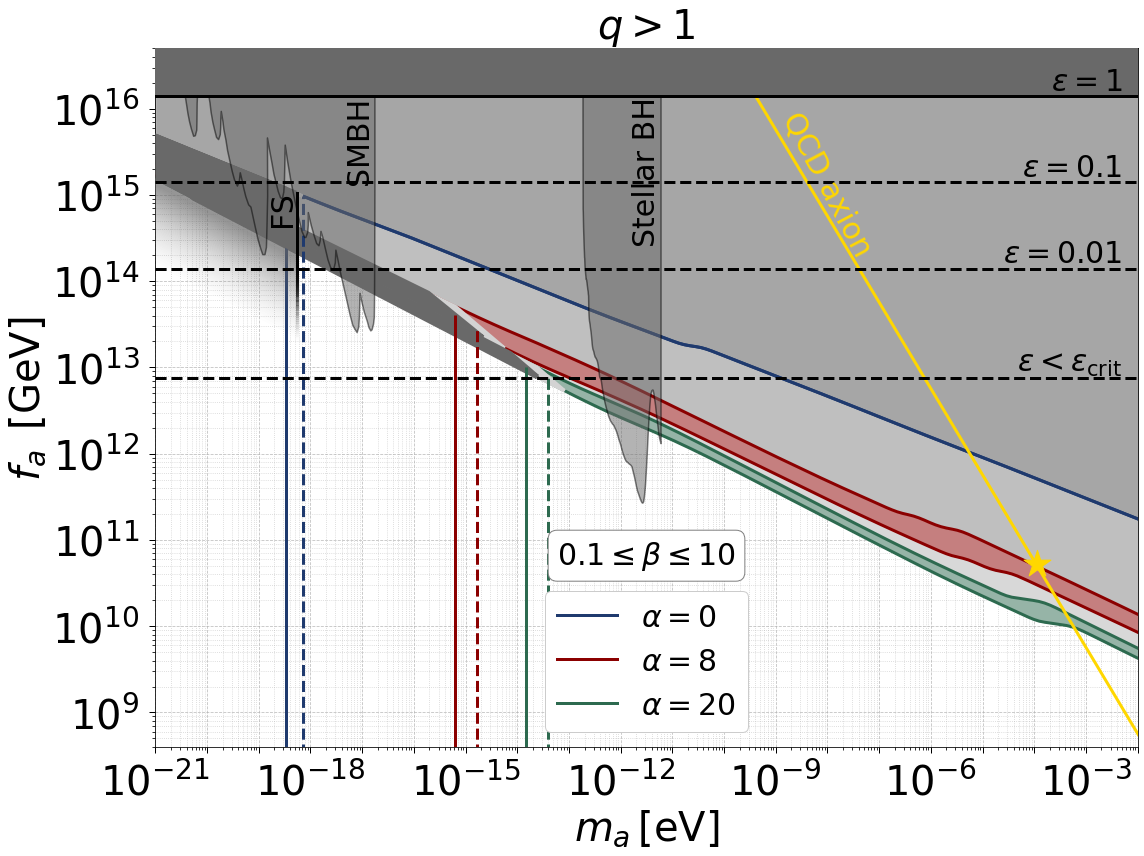}\quad
    \includegraphics[width=0.48\textwidth]{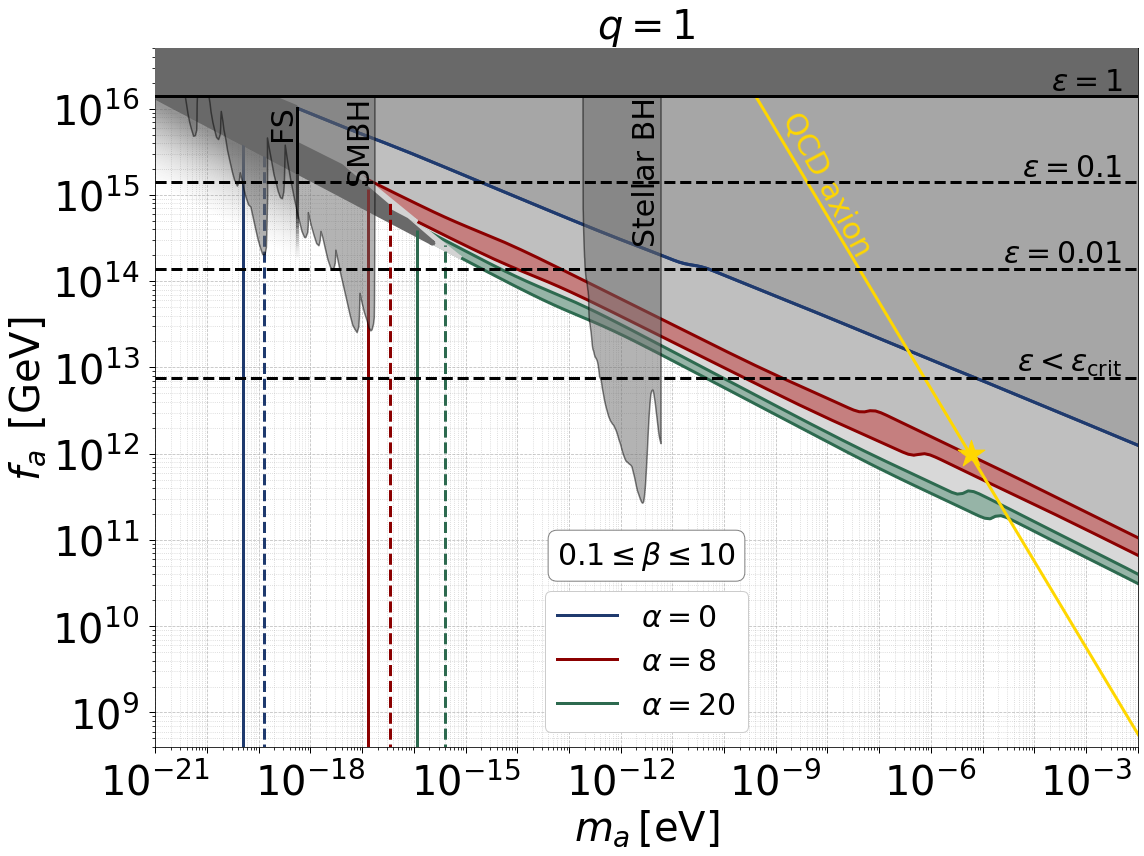}
    \caption{Same as \cref{fig:strings_abundance_q3} and \cref{fig:strings_abundance_q1} but for string density $\xi = 1$.}
    \label{fig:const_xi}
\end{figure}

We consider here the alternative possibility that the string network evolution follows exact scaling. In contrast to the logarithmic growth described in Eq.~(\ref{eq:xi}), this scenario assumes for the string density $\xi = 1$. The axion production from strings and the constraints are otherwise computed following the same procedure outlined in the main text. We present our results in \cref{fig:const_xi} in the same format. We show, as a function of the zero-temperature axion mass $m_a$, the value of $f_a$ required for the axion abundance produced by cosmic strings during the scaling regime to account for the observed dark matter density. The left panel corresponds to a spectral index $q>1$, while the right panel assumes a scale-free axion emission spectrum $q=1$. In both scenarios, the reduced string density leads to a smaller axion abundance, requiring larger values of $f_a$ to reproduce the observed dark matter density. In particular, for $q>1$ and a temperature-dependent axion mass, the SMBH superradiance constraints are no longer avoided, in contrast to the case with logarithmic violation of exact scaling. Even more significantly, the isocurvature constraints become significantly stronger, excluding axion masses up to $\sim10^{-15}$ eV for $q>1$ and up to $\sim10^{-16}$ eV for $q=1$ in the case of strong temperature dependence. 

%%%%%%%%%%%%%%%%%%%%%%%%%%%%%%%%%%%%%%%%%%%%%%%%%%%%%%%%%%%%%%%%
\section{Derivation of the power spectrum in the random phase model} \label{sec:random-phase-model}
%%%%%%%%%%%%%%%%%%%%%%%%%%%%%%%%%%%%%%%%%%%%%%%%%%%%%%%%%%%%%%%

In the random phase model, the axion field is expressed as a superposition of plane waves:
\begin{equation}
    a(t,\,\vec{x}) = \sum_{\vec{k}} a_{\vec{k}} e^{-i \left( \omega_{\vec{k}} t - \vec{k} \vec{x} \right)} e^{i \varphi_{\vec{k}}}\,,
\end{equation}
where the random phases $\varphi_{\vec{k}}$ are uniformly distributed in $[0,\,2\pi)$, ensuring statistical homogeneity and isotropy. The mode amplitudes $a_{\vec{k}}$ can be taken real and determine the contribution of each mode $\vec k$ to the overall field. The phases satisfy the following relations:
\begin{equation}\label{eq:phases}
  \langle e^{i \varphi_{\vec{k}}} \rangle = 0 \,,\quad
  \langle e^{i \left( \varphi_{\vec{k}} +  \varphi_{\vec{q}} \right) }\rangle = 0 \,,\quad
  \langle e^{i \left( \varphi_{\vec{k}} -  \varphi_{\vec{q}} \right) }\rangle = \delta_{\vec{k} \vec{q}} \,. 
\end{equation}
We consider the Fourier transform of the axion field in a large volume $V$:
\begin{equation}
\begin{aligned}
    \tilde{a}(t,\,\vec{k}) &= \int_{V} d^{3}x \, a(t,\,\vec{x}) e^{-i \vec{k} \vec{x}} 
    = V a_{\vec{k}} e^{-i \omega_{\vec{k}} t} e^{i \varphi_{\vec{k}}} \,,
\end{aligned}
\end{equation}
where we used 
\begin{equation}
    \int_{V} d^{3}x \, e^{-i \left( \vec{k} - \vec{k}^{\prime} \right) \vec{x}} = (2 \pi)^{3} \delta^{3}(\vec{k} - \vec{k}^{\prime}) 
\end{equation}
together with
\begin{equation}\label{eq:delta}
    \delta^{3}(\vec{k} - \vec{k}^{\prime}) = \frac{V}{(2 \pi)^{3}} \delta_{\vec{k} \vec{k}^{\prime}} \,.
\end{equation}
The normalization condition $\int d^3k\, \delta^3(\vec k - \vec k') = 1$ then implies the continuum limit 
\begin{equation}\label{eq:continuum}
 \int d^3 k \quad\longleftrightarrow\quad \frac{(2\pi)^3}{V} \sum_{\vec k} \,.
\end{equation}
Focusing on non-relativistic axions for simplicity, the local energy density is 
\begin{equation}\label{eq:rhoRPM}
\begin{aligned}
    \rho (x) & = m_{a}^{2} \lvert a(x)\rvert^{2} = 
    m_{a}^{2} \left[ \sum_{\vec{k}} a_{\vec{k}}^{2} +
      \sum_{\vec{k} \neq \vec{q}} a_{\vec{k}} a_{\vec{q}} \,
      e^{-i (\omega_{\vec{k}} - \omega_{\vec{q}}) t}
      e^{i (\vec{k} - \vec{q}) \vec{x}}
      e^{i (\varphi_{\vec{k}}- \varphi_{\vec{q}})} \right] \,.
\end{aligned}
\end{equation}
Averaging over random phases with \cref{eq:phases}, we see that only the first term contributes to the mean energy density
\begin{equation}
\begin{aligned}
  \bar{\rho} &= \langle m_{a}^{2} \lvert a(x)\rvert^{2} \rangle =
  m_{a}^{2} \sum_{\vec{k}} a_{\vec{k}}^{2} \,.
\end{aligned}
\end{equation}
Comparing this to \cref{eq:rho-bar} and using \cref{eq:continuum} gives the relation between the spectral density and the mode coefficients,
\begin{equation}\label{eq:coef}
  V m_a^2 a_{\vec k}^2 \quad\longleftrightarrow\quad \frac{2\pi^2}{k^2} \frac{\partial\rho}{\partial k} \,,
\end{equation}
which shows that $a_{\vec k}$ depends only on $k=|\vec k|$, consistent with isotropy.

The dimensionless density fluctuation field in the random phase model follows from the second term in \cref{eq:rhoRPM}:
\begin{equation}
\begin{aligned}
  \delta (x) \equiv \frac{\rho (x) - \bar{\rho}}{\bar{\rho}} =
\frac{m_{a}^{2}}{\bar\rho} 
      \sum_{\vec{k} \neq \vec{q}} a_{\vec{k}} a_{\vec{q}} \,
      e^{-i (\omega_{\vec{k}} - \omega_{\vec{q}}) t}
      e^{i (\vec{k} - \vec{q}) \vec{x}}
      e^{i (\varphi_{\vec{k}}- \varphi_{\vec{q}})} \,.
\end{aligned}
\end{equation}
Using again the random phase properties from \cref{eq:phases}, the two-point correlation function of the Fourier transform becomes
\begin{equation}
\begin{aligned}
    \langle \tilde{\delta} (k) \tilde{\delta}^{\ast} (k^{\prime}) \rangle    
    &=  V^2\frac{m_a^4}{\bar{\rho}^{2}} \sum_{\vec{l}}  a_{\vec{l}}^{2} a_{\vec{l} - \vec{k}}^{2} \delta_{\left( \vec{l} - \vec{k} \right) \left( \vec{l} - \vec{k}^{\prime} \right)} \\    
    &= \left( 2 \pi \right)^{3} V \frac{m_a^4} {\bar{\rho}^{2}} \sum_{\vec{l}}  a_{\vec{l}}^{2} a_{\vec{l} - \vec{k}}^{2} \delta^{3} \left( \vec{k} - \vec{k}^{\prime}\right) \,,
\end{aligned}
\end{equation}
where, in the last step, we used \cref{eq:delta} for the delta function. Comparing with the standard definition of the power spectrum,  $\langle \tilde{\delta}(\vec{k}) \tilde{\delta}^*(\vec{k}') \rangle = (2\pi)^3 P_\delta(k) \delta^3(\vec{k} - \vec{k}')$, we obtain
\begin{equation}
\begin{aligned}
  P_{\delta}(k) &= V  \frac{m_{a}^4}{\bar{\rho}^{2}} \sum_{\vec{l}}  a_{\vec{l}}^{2} a_{\vec{l} - \vec{k}}^{2}   \\
  &=\frac{\pi^2}{\bar\rho^2} \frac{1}{2\pi}\int\d^3l
  \, \frac{1}{l^2} \frac{\partial\rho}{\partial l}
  \, \frac{1}{{l'}^2} \frac{\partial\rho}{\partial l'}  
  \,,
\end{aligned}
\end{equation}
where, in the last step, we used \cref{eq:continuum,eq:coef} and defined
\begin{equation}
  l' \equiv |\vec l - \vec k| = \sqrt{l^2+k^2 - 2lk\cos\theta} \,.
\end{equation}
Switching to polar coordinates for the $d^3l$ integral, performing the trivial $\varphi$ integration and changing the integration variable from $d\cos\theta$ to $dl'$ reproduces \cref{eq:P_iso} in the main text.

%%%%%%%%%%%%%%%%%%%%%%%%%%%%%%%%%%%%%%%%%%%%%%%%%%%%%%%%%%%%%%%%
\section{Nonlinear transition due to axion self-interactions}
\label{app:self-int}

We briefly review the arguments provided in \cite{Gorghetto:2024vnp}, describing the second nonlinear transition in the case of $q>1,\,\alpha>0$. As discussed in \cref{sec:ALP_DM_post}, in this case the comoving axion number density is conserved after a first nonlinear transition at $T_\ell$. At that point, kinetic energy, gradient energy and potential energy are roughly equal: $m_\ell^2a_\ell^2 \sim (\nabla a_\ell)^2 \sim m_\ell^2 f_a^2$, where $a$ denotes the axion field. Hence, $a_\ell\sim f_a$. Let us now consider the potential energy from the axion self-interaction, emerging from the cosine potential in \cref{eq:V}, $V_{\rm self} \sim \lambda_a a^4$, with $\lambda_a(T) = m_a^2(T)/f_a^2$. Using $a_\ell \sim f_a$, we see that $V_{\rm self}(T_\ell) \sim m_\ell^2f_a^2$ is also of similar order as the other energy contributions. 

Consider now the evolution for $T_\ell > T > T_c$, where $T_c = \beta^{2/\alpha}\sqrt{m_af_a}$ is the temperature where the axion reaches its zero-temperature value. In this regime, the axion mass grows as $m_a(T) = m_\ell(T_\ell/T)^{\alpha/2}$. The conservation of comoving number density implies that $m_a(T)a^2(T) \propto T^3$, and therefore $a^2 \propto T^{3+\alpha/2}$. This leads to the following scaling:
\begin{align}
    m_a^2(T) a^2 \propto T^{3-\alpha/2} \,,\qquad
    (\nabla a)^2 \propto T^{5+\alpha/2} \,,\qquad
    \lambda_a a^4 \propto T^6 \qquad (T_\ell > T > T_c)\,.
\end{align}
We see that for $\alpha > 2$, the gradient energy drops faster than the self-energy, which leads to $(\nabla a)^2\ll \lambda_a a^4$ in this temperature regime. Physically, this implies that the system tends to contract under the attractive self-energy, until $(\nabla a)^2 \sim V_{\rm self}$ is re-established, which corresponds to a virialized state with a typical momentum $k_{\rm vir} \sim \sqrt{\lambda a^2}$. 
After the  axion has reached the zero-temperature mass, we have instead
\begin{align}
    m_a^2 a^2 \propto T^{3} \,,\qquad
    (\nabla a)^2 \propto T^{5} \,,\qquad
    \lambda_a a^4 \propto T^6 \qquad (T <T_c)\,,
\end{align}
which implies that $V_{\rm self}$ drops fastest and self-interactions can soon be neglected compared to gradient terms. 

The contraction towards virialization requires that low-momentum modes with $k<k_{\rm vir}$ are shifted towards $k_{\rm vir}$. This happens on a time scale \cite{Sikivie:2009qn,Gorghetto:2024vnp}
\begin{align}
    \tau_{\rm vir} \equiv \frac{1}{\Gamma_{\rm vir}} \sim \frac{8m_a(T)}{\lambda_a(T) a^2} \,.
\end{align}
If $\Gamma_{\rm vir} \gg H$, virialization takes place and the spectrum will peak at $k_{\rm peak} \sim k_{\rm vir}$. We find
\begin{align}\label{eq:GammaH}
    \frac{\Gamma_{\rm vir}}{H} \sim \frac{1}{8}\frac{m_\ell}{H_\ell}\frac{T}{T_\ell} \,,
\end{align}
which is maximal at $T=T_\ell$. Hence, if $\Gamma_{\rm vir} > H$ at $T_\ell$, virialization will start and last until either 
$\Gamma_{\rm vir}$ drops below $H$ or $T=T_c$, whatever happens earlier. Numerically, we find that for $q>1,\alpha\ge 8$, $\Gamma_{\rm vir} > H$ is fulfilled until $T_c$ and therefore, the spectrum will peak at 
\begin{align}
    k_{\rm peak} \sim k_{\rm vir}(T_c) \sim \sqrt{\lambda_a(T_c)a^2(T_c)} = \frac{m_a}{f_a}a(T_c) 
    \sim m_a \left(\frac{T_c}{T_\ell}\right)^{\frac{6+\alpha}{4}} \,.
\end{align}
This qualitative picture has been confirmed by numerical simulations in \cite{Gorghetto:2024vnp}, and motivates our assumption for the spectrum in \cref{eq:rho_DM_2nd-NL}.

In contrast, if $q=1$, we have $m_\ell\sim m_* = H_*$ and therefore \cref{eq:GammaH} implies that $\Gamma_{\rm vir}/H\sim 1/8 < 1$ already at $T_\ell\sim T_*$. This shows that for $q=1$ the nonlinear transition due to self-interactions is never relevant. Similarly, it is clear from this discussion that also for a temperature-independent axion mass, this effect does not occur. 

%%%%%%%%%%%%%%%%%%%%%%%%%%%%%%%%%%%%%%%%%%%%%%%%%%%%%%%%%%%%%%%

%%%%%%%%%%%%%%%%%%%%%%%%%%%%%%%%%%%%%%%%%%%%%%%%%%%%%%%%%%%%%%%%
\section{Isocurvature power spectrum in the pre-inflationary scenario} 
\label{app:iso_pre}
%%%%%%%%%%%%%%%%%%%%%%%%%%%%%%%%%%%%%%%%%%%%%%%%%%%%%%%%%%%%%%%

Assuming the axion is present during inflation with a field value $a$, the two-point correlation function of field perturbations sourced by quantum fluctuations related to the Gibbons-Hawking temperature of the inflation horizon, $T_{\rm GH} = H_I/(2\pi)$, is given by  
\begin{equation}
    \langle \delta a(\vec k) \delta a^*(\vec k') \rangle = \left (\dfrac{H_I}{2\pi} \right )^2 \dfrac{2\pi^2}{k^3}  
    \, V \delta_{\vec k{\vec k}'} \,,
\end{equation}
where $V$ is a finite volume, leading to discrete momenta. Since the energy density scales as $\rho \propto |a|^2$, the relative density fluctuation is $\delta \equiv \delta\rho/\rho = 2 \delta a/|a|$, yielding 
\begin{align}
    \langle \delta(\vec k)\delta^*(\vec k')\rangle = \frac{4}{|a|^2}
        \langle \delta a(\vec k) \delta a^*(\vec k') \rangle
        = \frac{2H_I^2}{k^3\theta_0^2f_a^2} 
            V \delta_{\vec k{\vec k}'} \,,
\end{align}
where, in the second step, we used $a=\theta_0f_a$ for the axion field value during inflation. Using the standard definition of the power spectrum, $\langle \delta(\vec k)\delta^*(\vec k')\rangle = (2\pi)^3\delta^3(\vec k - \vec k') P_\delta(k)$, and applying the continuum limit $V \delta_{\vec k{\vec k}'} \to 
(2\pi)^3\delta^3(\vec k - \vec k')$ (see \cref{sec:random-phase-model}), we obtain for the dimensionless power spectrum
$\Delta^2_a(k) \equiv P_\delta(k) k^3/(2\pi^2)$ the expression given in \cref{eq:Delta_pre}.

\acknowledgments
We thank Marco Gorghetto, Ed Hardy, Mathieu Kaltschmidt and Javier Redondo for useful discussions and comments on the manuscript.

\bibliographystyle{JHEP}
\bibliography{./biblio}
\end{document}